%% file: kit-TCOM2020-FAMAv2-sc.tex
\documentclass[11pt]{article}

\usepackage{sectsty}
\sectionfont{\sc}
\renewcommand\thesection{\Roman{section}.}
\renewcommand\thesubsection{\Alph{subsection}.}

\usepackage{graphicx}
\usepackage{citesort}
\usepackage{amssymb}
\usepackage{amsfonts}
\usepackage{amsmath}
\usepackage{epsfig}
\usepackage{fancybox}
\usepackage{textcomp}
\usepackage{multirow}
\usepackage{appendix}
\usepackage{setspace}
\usepackage{subfigure}

\usepackage[top=1.3cm,bottom=1.3cm,left=1.5cm,right=1.5cm]{geometry}
\renewcommand{\baselinestretch}{1.75}
\input macro

\newtheorem{Corollary}{Corollary}

\begin{document}

\title{\Huge\bf Fluid Antenna Multiple Access\thanks{The work is supported by EPSRC under grant EP/M016005/1.}}

\setcounter{footnote}{2}

\author{Kai-Kit Wong\thanks{Department of Electronic and Electrical Engineering, University College London, London, United Kingdom.}, and Kin-Fai Tong$^\S$}

\maketitle
\thispagestyle{empty}

\begin{abstract}
Fluid antenna is a novel technology that can make an antenna appear instantly at one of $N$ preset locations in a predefined space. An important application is to adopt fluid antenna in a small space of mobile device for obtaining the tremendous diversity hidden in the small space. Previous results have revealed that a single-antenna fluid antenna system, even with a very small space, can outperform a multiple antenna maximum ratio combining (MRC) system if $N$ is large enough. This paper explores the potential of using fluid antenna for multiple access through performance analysis. Fluid antenna multiple access (FAMA) exploits moments of deep fade experienced by the interference to achieve a favourable channel condition for the desired signal, without requiring sophisticated signal processing. We analyze the FAMA system by first deriving the outage probability of the signal-to-interference ratio (SIR) in a double integral form. We then obtain an outage probability upper bound in closed form and an average outage capacity lower bound for the FAMA system, with an arbitrary number of interferers, from which the multiplexing gain of FAMA is characterized. We also estimate how large $N$ is required to achieve a given  multiplexing gain using fluid antennas with a given size. Results illustrate that it is possible for FAMA to support hundreds of users using only one fluid antenna at each user in a few wavelengths of space, giving rise to significant enhancement in the network outage capacity.

\begin{center}
{\bf Index Terms}
\end{center}
FAMA, Fluid antennas, MIMO, Multiple access, Selection combining, Outage.
\end{abstract}

\thanks{\singlespacing}
\setcounter{page}{0}

\newpage
\section{Introduction}
Over the past twenty years, multiple-input multiple-output (MIMO) has revolutionized mobile communications by creating bandwidth from space, independent of frequency and time resources. The enormous diversity and multiplexing gains inherent in spatially uncorrelated channels at multiple antennas have time and time again proven to deliver extraordinary performance for point-to-point communication links, e.g., \cite{Paulraj-1994,Foschini-1998,Varokh-1998,Alamouti-1998,Tse-2003}. Multiuser MIMO, however, is arguably an even greater contribution, as it enables us to multiplex users entirely in the spatial domain and scales up the network capacity with the number of antennas at the base stations (BSs) and/or mobile stations (MSs) without co-channel interference, e.g., \cite{Goldsmith-2003,Spencer-2004,Marzetta-2013,Marzetta-2014}. 

Multiuser MIMO focuses predominantly on the signal processing and coding at the BS side, leaving the MSs a relatively easier task to achieve communications. This makes sense because BSs tend to have much higher processing capability but it also means that the BSs could quickly become overwhelmed by the processing requirements of a huge number of users. Although massive MIMO greatly simplifies the signal processing of multiuser signals using the law of large numbers by adopting a massive number of antennas at the BS \cite{Marzetta-2013,Marzetta-2014}, the number of antennas used needs to be much greater than the number of supported users. Also, for 5G as an example, only $64$ antennas are employed and channel inverses are required to eliminate inter-user interference \cite{Larsson-2019,Sprint-2019,Huawei-2019}, let alone other associated processing tasks such as pilot decontamination \cite{Yin-2016}, power control \cite{Cheng-2017}, and user-cell association \cite{Caire-2016}, etc. The fact that the latest trends require resource management to be service-based, content-centric \cite{Tao-2020,You-2020}, further puts strains on the BSs.

\begin{quote}
{\em Can space-division multiple-access be much simpler, at least conceptually?}
\end{quote}

This paper attempts to provide a positive argument to answer this question. We begin by reviewing the phenomenon of fading. Multipath fading, also known as short-term fading, is a result of radio waves travelling via different paths due to scattering from a transmitter to a receiver. The complex interaction between the paths and the physical environment gives rise to a mixture of construction or destruction of multipath signals, resulting in a random signal envelope at the receiver. When the signal goes into a deep fade, as deep as $50$dB drop in the field strength is not uncommon. A great deal of research in the past has been conducted to avoid these deep fades. By contrast, this paper aims to exploit those moments of deep fade for multiple access, and deep fades of $50$dB, if occurring for the interference, are desirable. 

The technology that enables us to skim through a collection of fading envelopes and takes advantage of their ups and downs is fluid antennas \cite{Wong-2020twc,Wong-2020cl}. Fluid antenna is an emerging concept that describes an apparatus that has the ability to instantly switch the position of an antenna to one of $N$ preset locations (referred to as `ports') within a small space. The concept is greatly motivated by the recent advances in mechanically flexible antennas such as liquid metal antennas \cite{Hayes-2012,Ohta-2013,Dey-2016,Soh-2020} or other variants \cite{Saghati-2014,Tong-2017,Tong-2018,Singh-2019,Tong-2019}, as well as pixel-like antennas \cite{Murch-2014}. In \cite{Wong-2020twc,Wong-2020cl}, a fluid antenna system with a mechanically flexible antenna\footnote{In this paper, being `mechanically flexible' does not necessarily mean that the antenna is physically mobilized from one position to another because activating or deactivating radiating elements at different positions may be achieved by electronically controlled pixels. However, the concept of fluid antenna may be easily visualized as an antenna with a flexible position.} over a small linear space was considered, and it was reported that even with a tiny space of half-wavelength or less, the fluid antenna system can deliver capacity that is achievable by a multi-antenna maximum ratio combining (MRC) system, all by a single fluid antenna with a single RF chain, if $N$ is large enough. Spatial correlation in a small space being the limitation therefore may need to be reconsidered.

\begin{figure}[h]
\begin{center}
\includegraphics[width=17cm]{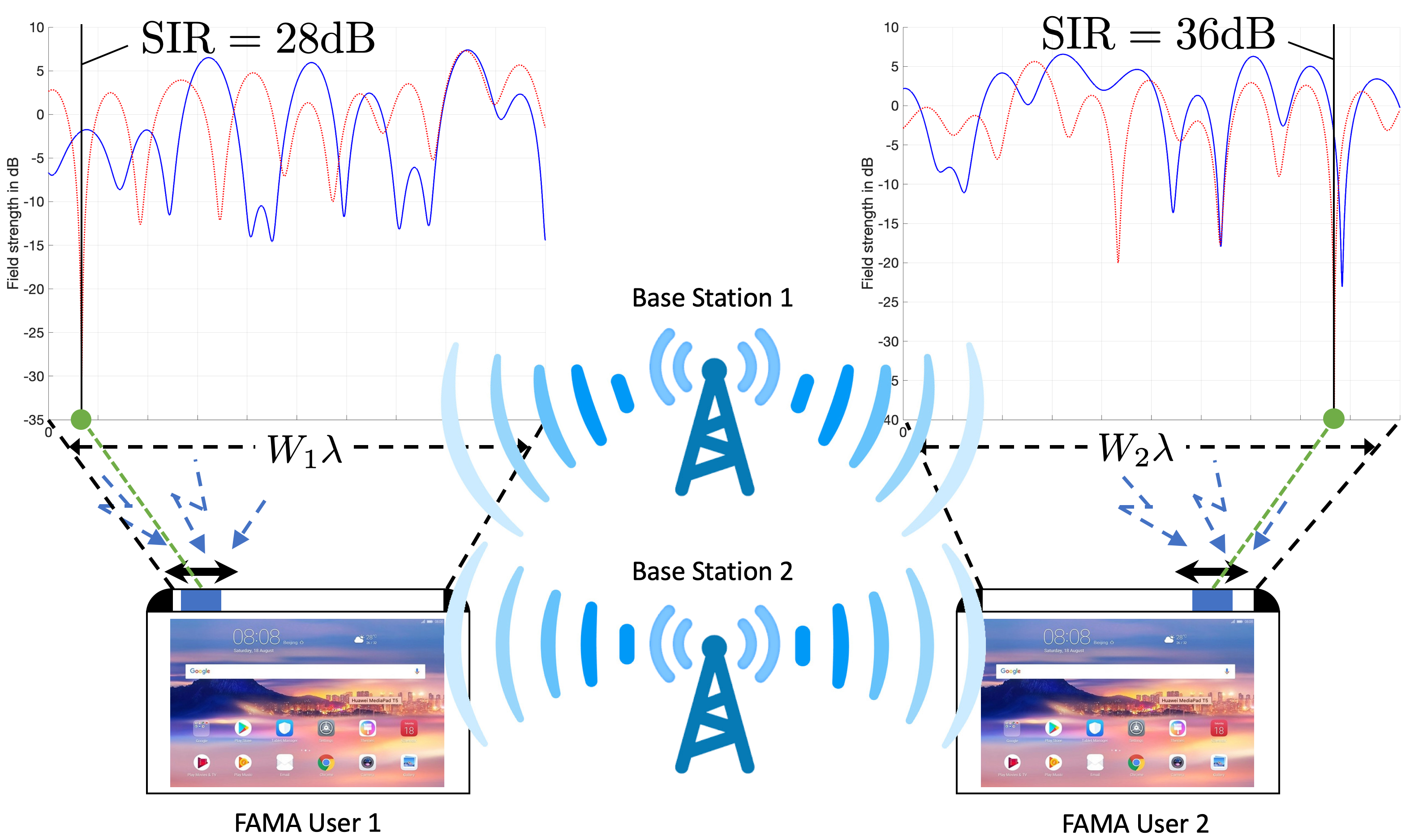}
\caption{The concept of FAMA where each FAMA user is able to switch its antenna to the position in which the interference is in a deep fade for maximizing the SIR. FAMA works for interference channels as the channels for the desired signal and interference have different fading envelopes.}\label{fig:fama}
\end{center}
\end{figure}

In this paper, we extend the use of fluid antennas for multiuser communications. In particular, we consider an interference channel where there are multiple pairs of transmitters and receivers. Figure \ref{fig:fama} illustrates a possible scenario with two transmitter-receiver pairs, and that both MSs (or users) deploy a fluid antenna system for reception. The fluid antenna system for user 1 with a space of $W_1\lambda$ where $\lambda$ is the wavelength of communication, is able to observe the fading envelopes in the available space and can switch the antenna to the position where the signal for user 2 is in a deep fade. Same happens for user 2 when being optimized to mitigate the interference from BS transmitter 1. Remarkably, in this case, no fancy signal processing is required but inter-user interference is naturally avoided by picking the right moment (in space) where the signal-to-interference ratio (SIR) is maximized for multiple access.

In this paper, we refer to this approach as fluid antenna multiple access (FAMA) where each MS (or a FAMA user) has an $N$-port fluid antenna to always switch to the position with the strongest SIR. FAMA is entirely user-centric and requires no coordination between transmitters and receivers. The processing at a FAMA user also has no impact on other coexisting users, greatly simplifying resource management. The interference suppression capability comes naturally from the fading phenomenon, and the degree of freedom is determined by the richness of scattering in the environment, and the operational parameters such as the number of ports, $N$, and the size of the fluid antenna, $W\lambda$, at each user.

The aim of this paper is to characterize the achievable performance of FAMA and understand how the performance scales with the number of ports and the size of the fluid antenna at each user. In particular, one objective is to derive the outage probability for the SIR which reveals the interference suppression capability at the user level. The exact outage probability expression for the SIR is obtained in a double integral form. Then we propose an outage probability upper bound in closed form which will permit us to unravel the relationship between the operational parameters and the achievable performance. A lower bound for the average outage capacity for the FAMA network is also presented, which is followed by our proposed definition of multiplexing gain to measure the capacity increase by FAMA. We also derive a sufficient condition on how large $N$ is required for FAMA to achieve a certain multiplexing gain.

In addition to the above technical contributions, we highlight some key findings as follows:
\begin{itemize}
\item Space of the fluid antenna at the MS is important but has a diminishing return after $\frac{\lambda}{2}$, and its impact on the network outage capacity is insignificant if the size grows beyond $\frac{\lambda}{2}$.\footnote{That said, space can be extremely important as $N$ is anticipated to be constrained by space for implementation.}
\item If the number of ports, $N$, is sufficiently large, then a single fluid antenna at the MS can achieve any arbitrarily small SIR outage probability, demonstrating the feasibility of FAMA. 
\item The network outage capacity scales linearly with the number of ports at each FAMA user but is ultimately limited by the number of coexisting users (or MSs).
\item The multiplexing gain of FAMA scales linearly with the number of ports, and is inversely proportional to the SIR target but is again upper bounded by the number of coexisting MSs.
\end{itemize}

The rest of the paper is organized as follows. In Section II, we introduce the network model of FAMA under the setting of an interference channel. Our main results, which focus on the outage probability analysis of SIR at a fluid antenna system and the outage capacity of the FAMA network, are presented in Section III. Section IV attempts to make some interesting observations of FAMA by presenting some numerical results. Finally, we provide some concluding remarks in Section V.


\section{Network Model}\label{sec:model}
\subsection{Single-User Fluid Antenna System}\label{ssec:sufas}
Before we consider a network of fluid antenna system users, we first introduce the model for a single MS equipped with a fluid antenna. The MS operates as a receiver and has a `fluid' antenna whose location can be switched instantly to one of the $N$ preset locations evenly distributed along a linear dimension of length, $W\lambda$, where $\lambda$ is the wavelength of communication. We refer to a switchable location as `port' and all the ports share a common RF chain. Any delay for switching between the ports is assumed negligible, and ignored in this paper. Moreover, in order to make possible performance analysis using communication theory, we use an abstraction to model the concept and treat each port as an ideal point antenna.  

The first port is the reference location, from which the displacement of the $k$-th port is measured:
\begin{equation}\label{eqn:dn}
d_k=\left(\frac{k-1}{N-1}\right)W\lambda,~\mbox{for }k=1,2,\dots,N.
\end{equation}
The received signal at the $k$-th port is modelled as
\begin{equation}\label{eqn:yk}
z_k=g_k s+\eta_k,
\end{equation}
where the time index is omitted for conciseness, $g_k$ is the complex channel coefficient experienced by the $k$-th port, which follows a circularly symmetric complex Gaussian distribution with zero mean and variance of $\sigma^2$, $\eta_k$ denotes the complex additive white Gaussian noise (AWGN) at the $k$-th port with zero mean and variance of $\sigma_\eta^2$, and $s$ denotes the transmitted data symbol. In our model, the amplitude of the channel, $|g_k|$, is assumed Rayleigh distributed, with the probability density function (pdf)
\begin{equation}\label{eqn:|g|}
p_{|g_k|}(r)=\frac{2r}{\sigma^2}e^{-\frac{r^2}{\sigma^2}},~\mbox{for }r\ge 0~\mbox{with }{\rm E}[|g_k|^2]=\sigma^2.
\end{equation}
The average received signal-to-noise ratio (SNR) at each port is given by
\begin{equation}\label{eqn:snr}
\Gamma=\sigma^2\frac{{\rm E}[|s|^2]}{\sigma^2_\eta}\equiv\sigma^2\Theta,~\mbox{where }\Theta\triangleq\frac{{\rm E}[|s|^2]}{\sigma^2_\eta}.
\end{equation}
The channels $\{g_k\}_{\forall n}$ are considered to be correlated as they can be arbitrarily close to each other. 

With 2-D isotropic scattering and an isotropic receiver port, it is known that the autocorrelation functions of the channel satisfy \cite{Wong-2020twc,Stuber-2002}
\begin{equation}\label{eqn:space-correlation}
\phi_{g_kg_\ell}(d_k-d_{\ell})=\frac{\sigma^2}{2}J_0\left(2\pi\frac{d_k-d_{\ell}}{\lambda}\right)=\frac{\sigma^2}{2}J_0\left(\frac{2\pi(k-\ell)}{N-1}W\right),
\end{equation}
where $J_0(\cdot)$ denotes the zero-order Bessel function of the first kind. For ease of exposition, we find it useful to parameterize the channels at the $N$ antenna ports by 
\begin{equation}\label{eqn:g-model}
\left\{\begin{aligned}
g_1&=\sigma x_0 +j\sigma y_0\\
g_k&=\sigma\left(\sqrt{1-\mu_k^2}x_k+\mu_k x_0\right)+j\sigma\left(\sqrt{1-\mu_k^2}y_k+\mu_k y_0\right),~\mbox{for }k=2,\dots,N,
\end{aligned}\right.
\end{equation}
where $x_0,x_1,\dots,x_N,y_0,y_1,\dots,y_N$ are all independent Gaussian random variables with zero mean and variance of $\frac{1}{2}$, and $\{\mu_k\}$ are the autocorrelation parameters that can be chosen appropriately to specify the correlation among $\{g_k\}$. Based on this model, ${\rm E}[|g_k|^2]=\sigma^2$ for all $k$ and due to (\ref{eqn:space-correlation}), we have
\begin{equation}\label{eqn:mu-condition}
\mu_k=J_0\left(\frac{2\pi(k-1)}{N-1}W\right),~\mbox{for }k=2,\dots,N.
\end{equation}

\subsection{FAMA}\label{ssec:fama}
Now, we consider that there are $N_{\rm I}$ interferers (i.e., $N_{\rm I}+1$ users). Therefore, (\ref{eqn:yk}) becomes
\begin{equation}\label{eqn:yk-fama}
z_k=g_k s+\sum_{i=1}^{N_{\rm I}}g_k^{(i)}s_i+\eta_k\equiv g_k s+g_k^{\rm I}+\eta_k,
\end{equation}
where $s_i$ is the transmitted data from the $i$-th interferer and $g_k^{(i)}$ denotes its corresponding channel. The parameters for the interferers are modelled in the same way as the desired signal. Since $\{g_k^{(i)}\}_{\forall i}$ are all complex Gaussian, we can model the total interference at the $k$-th port as $g^{\rm I}_k=\sum_{i=1}^{N_{\rm I}}g_k^{(i)}s_i$ which is again complex Gaussian with zero mean and variance of some $\sigma_{\rm I}^2=\sum_{i=1}^{N_{\rm I}}{\rm E}[|g_k^{(i)}|^2]{\rm E}[|s_i|^2]$.

Following from (\ref{eqn:g-model}), we can also model $\{g_k^{\rm I}\}$ by
\begin{equation}\label{eqn:gi-model}
\left\{\begin{aligned}
g_1^{\rm I}&=\sigma_{\rm I} x_0^{\rm I} +j\sigma_{\rm I} y_0^{\rm I}\\
g_k^{\rm I}&=\sigma_{\rm I}\left(\sqrt{1-\mu_k^2}x_k^{\rm I}+\mu_k x_0^{\rm I}\right)+j\sigma_{\rm I}\left(\sqrt{1-\mu_k^2}y_k^{\rm I}+\mu_k y_0^{\rm I}\right),~\mbox{for }k=2,\dots,N,
\end{aligned}\right.
\end{equation}
where $x_0^{\rm I},x_1^{\rm I},\dots,x_N^{\rm I},y_0^{\rm I},y_1^{\rm I},\dots,y_N^{\rm I}$ are all independent Gaussian random variables with zero mean and variance of $\frac{1}{2}$, and $\{\mu_k\}$ are the autocorrelation parameters satisfying (\ref{eqn:mu-condition}). Note that our model focuses on a typical FAMA user in the presence of interference and omits the user index because the model applies to all users in the same way which have the same average performance if their statistics are identical.

At this FAMA user, it is assumed that the fluid antenna can always switch to the maximum of $\frac{|g_k|}{|g_k^{\rm I}|}$ in order to have the best reception performance. Therefore, we are interested in the random variable
\begin{equation}
g_{\rm FAMA}=\max\left\{\frac{|g_1|}{|g_1^{\rm I}|},\frac{|g_2|}{|g_2^{\rm I}|},\dots,\frac{|g_N|}{|g_N^{\rm I}|}\right\}.
\end{equation}
The channels for the desired signal $\{|g_k|\}$ and that for the interference $\{|g_k^{\rm I}|\}$ are assumed independent but they are correlated random variables across space, specified by (\ref{eqn:mu-condition}), which causes their ratios to be correlated in a particular way. The random variable $g_{\rm FAMA}$ corresponds to the square root of the SIR of the received signal. Although signal-to-interference plus noise ratio (SINR) is a more accurate performance measure, SIR is chosen to ease the analysis. In addition, for interference-limited environments where $N_{\rm I}$ and/or $\sigma_{\rm I}$ are large, the use of SIR represents a judicious approximation.


\section{Main Results}
In this section, we present our analysis on the achievable performance of FAMA. We will start by presenting the outage probability for the SIR of a FAMA user, and a closed-form outage probability upper bound. Then we will introduce an average outage capacity lower bound which we use to measure the overall outage capacity performance of a FAMA network. A definition of multiplexing gain will then be given, from which we will investigate how the system parameters such as the number of ports and size of the fluid antenna impact on the multiplexing gain and capacity performance of the network.

\begin{theorem}\label{th:op-fama}
The outage probability of the SIR, with a target $\gamma$, for a FAMA user is given by
\begin{multline}\label{eqn:op-sir}
{\rm Prob}\left({\rm SIR}\le\gamma\right)=\int_0^\infty e^{-z}\int_0^{\frac{\sigma_{\rm I}^2\gamma}{\sigma^2}z}e^{-t}\left\{\prod_{k=2}^N\left[1+\left(\frac{\frac{\sigma_{\rm I}^2\gamma}{\sigma^2}}{\frac{\sigma_{\rm I}^2\gamma}{\sigma^2}+1}\right)e^{-\left(\frac{1}{\frac{\sigma_{\rm I}^2\gamma}{\sigma^2}+1}\right)\frac{\mu_k^2}{1-\mu_k^2}\left(\frac{\sigma_{\rm I}^2\gamma}{\sigma^2}z+t\right)}\times\right.\right.\\
\left.\left.I_0\left(\frac{\frac{\sigma_{\rm I}\sqrt{\gamma}}{\sigma}}{\frac{\sigma_{\rm I}^2\gamma}{\sigma^2}+1}\left(\frac{2\mu_k^2}{1-\mu_k^2}\right)\sqrt{zt}\right)-Q_1\left(\frac{1}{\sqrt{\frac{\sigma_{\rm I}^2\gamma}{\sigma^2}+1}}\sqrt{\frac{2\mu_k^2}{1-\mu_k^2}}\sqrt{t},\sqrt{\frac{\frac{\sigma_{\rm I}^2\gamma}{\sigma^2}}{\frac{\sigma_{\rm I}^2\gamma}{\sigma^2}+1}}\sqrt{\frac{2\mu_k^2}{1-\mu_k^2}}\sqrt{z}\right)\right]\right\} dt dz,
\end{multline}
where $I_0(\cdot)$ is the zero-order modified Bessel function of the first kind, and $Q_1(\cdot,\cdot)$ denotes the first-order Marcum-$Q$ function. The parameters $\sigma$, $\sigma_{\rm I}$ and $\{\mu_k\}$ are defined in the previous section.
\end{theorem}

\proof See Appendix A.
\endproof

Theorem \ref{th:op-fama} gives out the exact outage probability expression for the SIR at a FAMA user. However, if $z\to\infty$, $I_0(z)\to\infty$, which happens when evaluating the integral (\ref{eqn:op-sir}). To get around this, the second term inside the product can be evaluated by
\begin{multline}
\left(\frac{\frac{\sigma_{\rm I}^2\gamma}{\sigma^2}}{\frac{\sigma_{\rm I}^2\gamma}{\sigma^2}+1}\right)e^{-\left(\frac{1}{\frac{\sigma_{\rm I}^2\gamma}{\sigma^2}+1}\right)\frac{\mu_k^2}{1-\mu_k^2}\left(\frac{\sigma_{\rm I}^2\gamma z}{\sigma^2}+t\right)}I_0\left(\frac{\frac{\sigma_{\rm I}\sqrt{\gamma}}{\sigma}}{\frac{\sigma_{\rm I}^2\gamma}{\sigma^2}+1}\left(\frac{2\mu_k^2}{1-\mu_k^2}\right)\sqrt{zt}\right)\\
=\frac{1}{\pi}\left(\frac{\frac{\sigma_{\rm I}^2\gamma}{\sigma^2}}{\frac{\sigma_{\rm I}^2\gamma}{\sigma^2}+1}\right)
e^{-\left(\frac{1}{\frac{\sigma_{\rm I}^2\gamma}{\sigma^2}+1}\right)\frac{\mu_k^2}{1-\mu_k^2}\left(\frac{\sigma_{\rm I}\sqrt{\gamma z}}{\sigma}-\sqrt{t}\right)^2}\int_0^\pi e^{-\left(\frac{1}{\frac{\sigma_{\rm I}^2\gamma}{\sigma^2}+1}\right)\frac{\mu_k^2}{1-\mu_k^2}\frac{4\sigma_{\rm I}\sqrt{\gamma zt}}{\sigma}\sin^2\frac{\theta}{2}}d\theta.
\end{multline}
The above result can be easily obtained from the definition of $I_0(\cdot)$.

Although numerical integration for (\ref{eqn:op-sir}) is possible, the computational complexity when $N$ is large is very high, and the integral form also does not allow any insight to be gained. We therefore develop an outage probability upper bound, which we present in the next theorem.

\begin{theorem}\label{th:op-ub1}
The SIR outage probability in (\ref{eqn:op-sir}) is upper bounded by
\begin{align}
{\rm Prob}\left({\rm SIR}\le\gamma\right)&\le\varepsilon_{\rm UB}^{\rm I}(\gamma)\notag\\
&=\int_0^\infty e^{-z}\left(1-e^{-\frac{\sigma_{\rm I}^2\gamma}{\sigma^2}z}\right)\prod_{k=2}^N\left[1-\left(\frac{1}{\frac{\sigma_{\rm I}^2\gamma}{\sigma^2}+1}\right)\frac{e^{-\left(\frac{1}{\frac{\sigma_{\rm I}^2\gamma}{\sigma^2}+1}\right)\left(\frac{\mu_k^2}{1-\mu_k^2}\right)\frac{\sigma_{\rm I}^2\gamma}{\sigma^2}z}}{1+4\left(\frac{1}{\frac{\sigma_{\rm I}^2\gamma}{\sigma^2}+1}\right)\left(\frac{\mu_k^2}{1-\mu_k^2}\right)\frac{\sigma_{\rm I}^2\gamma}{\sigma^2}z}\right]dz.\label{eqn:op-ub1}
\end{align}
\end{theorem}

\proof See Appendix B.
\endproof

\begin{Corollary}\label{co:sir2zero}
As $N\to\infty$, the SIR outage probability goes to $0$ as long as $|\mu_k|\ne 1~\forall k$.
\end{Corollary}

\proof It can be seen from $(\ref{eqn:op-ub1})$ that the integrand is a product of $N+1$ less-than-one numbers. If $N\to\infty$, it therefore goes to $0$ which completes the proof.
\endproof

Theorem \ref{th:op-ub1} appears to identify the benefit of each individual port in reducing the outage probability and Corollary \ref{co:sir2zero} confirms the feasibility of FAMA since the interference at the MS can always be avoided if $N$ is large enough. Also, (\ref{eqn:op-ub1}) further allows a closed-form outage probability upper bound to be derived for the special case where $|\mu_2|=\cdots=|\mu_N|=\mu$. This result is presented in the following theorem.

\begin{theorem}\label{th:op-ub2}
If $|\mu_2|=\cdots=|\mu_N|=\mu$, then the SIR outage probability is upper bounded by
\begin{equation}\label{eqn:op-ub2}
{\rm Prob}\left({\rm SIR}\le\gamma\right)\lesssim\varepsilon_{\rm UB}^{\rm II}(\gamma)
=\frac{1-\mu^2}{4\mu^2}e^{\frac{1-\mu^2}{4\mu^2}}\sum_{k=0}^{N-1}\frac{{N-1 \choose k}(-1)^k}{\left(\frac{\sigma_{\rm I}^2\gamma}{\sigma^2}\right)^k}e^{\frac{k}{4}}E_k\left(\frac{k}{4}+\frac{1-\mu^2}{4\mu^2}\right),
\end{equation}
where $E_k(\cdot)$ denotes the generalized exponential integral.
\end{theorem}

\proof See Appendix C.
\endproof

The following theorem quantifies the network capacity performance of FAMA.

\begin{theorem}\label{th:cap-fama}
The average network outage capacity of FAMA is lower bounded by
\begin{align}
C_{\rm FAMA}(\gamma)
&\approx(N_{\rm I}+1)\left(1-{\rm Prob}({\rm SIR}\le\gamma)\right)\log_2(1+\gamma)\notag\\
&\ge (N_{\rm I}+1)\left(1-\varepsilon_{\rm UB}(\gamma)\right)\log_2(1+\gamma),\label{eqn:C-fama}
\end{align}
where $\varepsilon_{\rm UB}(\gamma)$ can be either $\varepsilon_{\rm UB}^{\rm I}(\gamma)$ in (\ref{eqn:op-ub1}) or $\varepsilon_{\rm UB}^{\rm II}(\gamma)$ in (\ref{eqn:op-ub2}).
\end{theorem}

\proof For each FAMA user, the $\theta$-outage capacity is defined as 
\begin{equation}
C_{\rm out}^\theta(\gamma)=\log_2(1+\gamma)
\end{equation}
such that ${\rm Prob}({\rm SINR}\le\gamma)=\theta$. As the transmitter does not know when outage occurs, it transmits its information at a rate of $\log_2(1+\gamma)$, and therefore, the average outage capacity becomes $(1-\theta)\log_2(1+\gamma)$. For the FAMA network, since all the users are independent, the average outage capacity of the whole network therefore is the average outage capacity for a typical FAMA user scaled by the total number of users. Replacing the exact outage probability by an upper bound finally gives (\ref{eqn:C-fama}).
\endproof

To measure the capacity benefit of FAMA, we introduce the following definition of multiplexing gain.

\begin{definition}
The multiplexing gain for the FAMA network is defined as
\begin{equation}\label{eqn:mg}
m\triangleq\left(N_{\rm I}+1\right)\left(1-\varepsilon_{\rm UB}(\gamma)\right).
\end{equation}
\end{definition}

The above definition basically interprets the outage capacity lower bound of the FAMA network as the capacity of a single-user communication link achieving an SNR of $\gamma$, scaled by the multiplexing gain $m$. Note that normally, if we attempt to overlap users without any interference avoidance mechanism, $\varepsilon_{\rm UB}(\gamma)$ will be close to one for any meaningful $\gamma$, and $m\approx 0$, which is worse than single-user communications. In FAMA, on the contrary, $\varepsilon_{\rm UB}(\gamma)$ is reduced to an acceptable level by switching to the port where the interference undergoes a deep fade, achieving the maximum SIR. This is more possible when $N$ and/or $W$ are large. We will analyze how those system parameters affect the capacity performance below.

\begin{theorem}\label{th:Nvsm}
FAMA achieves a multiplexing gain of $m$ or more with a target SIR $\gamma$ if $N$ satisfies
\begin{equation}\label{eqn:Ncon-1}
\frac{1-\mu^2}{4\mu^2}e^{\frac{1-\mu^2}{4\mu^2}}\sum_{k=0}^{N-1}\frac{{N-1 \choose k}(-1)^k}{\left(\frac{\sigma_{\rm I}^2\gamma}{\sigma^2}\right)^k}e^{\frac{k}{4}}E_k\left(\frac{k}{4}+\frac{1-\mu^2}{4\mu^2}\right)\le 1-\frac{m}{N_{\rm I}+1},
\end{equation}
which, under the condition that $\mu$ is reasonably small, further yields the simplified condition
\begin{equation}\label{eqn:Ncon-2}
\sum_{k=1}^{N-1}\frac{{N-1 \choose k}(-1)^{k+1}}{\left(\frac{\sigma_{\rm I}^2\gamma}{\sigma^2}\right)^k}\left(1-k\mu^2\right)\gtrsim\frac{m}{N_{\rm I}+1}.
\end{equation}
For an ambitious target $\gamma$ and very large interference power $\sigma_{\rm I}^2$, i.e., $\frac{\sigma_{\rm I}^2\gamma}{\sigma^2}$ is very large, we have
\begin{equation}\label{eqn:Ncon-3}
N\gtrsim \left(\frac{m}{N_{\rm I}+1}\right)\frac{\left(\frac{\sigma_{\rm I}^2\gamma}{\sigma^2}\right)}{1-\mu^2}+1\approx\frac{m\gamma}{1-\mu^2}+1,
\end{equation}
where the last result arises when $\sigma_{\rm I}^2=N_{\rm I}\sigma^2$ and $N_{\rm I}$ is large.
\end{theorem}

\proof See Appendix D.
\endproof

Note that the power of the total interference $\sigma_{\rm I}^2$ can be linked to the number of interferers occupying in the same environment. If all the interferers have the same power and that their power is also identical to that of the desired user, then $\sigma_{\rm I}^2=N_{\rm I}\sigma^2$, and hence $\frac{\sigma_{\rm I}^2\gamma}{\sigma^2}=N_{\rm I}\gamma$. The condition (\ref{eqn:Ncon-3}) makes sense as $N$ appears to be directly proportional to the multiplexing gain and the target SIR. In addition, high autocorrection $\mu$ will require a larger $N$ to compensate for the performance. The next corollary explicitly illustrates how the achievable multiplexing gain scales with the system parameters.

\begin{Corollary}
The multiplexing gain of the FAMA network is given by
\begin{equation}\label{eqn:m-FAMA}
m\approx\min\left\{\frac{(N-1)(1-\mu^2)(N_{\rm I}+1)}{\left(\frac{\sigma_{\rm I}^2\gamma}{\sigma^2}\right)},N_{\rm I}+1\right\}
\approx\min\left\{\frac{(N-1)(1-\mu^2)}{\gamma},N_{\rm I}+1\right\}.
\end{equation}
\end{Corollary}

\proof This comes directly from (\ref{eqn:Ncon-3}) and (\ref{eqn:mg}).
\endproof

On the other hand, in terms of interference immunity or the SIR outage probability, a FAMA network serving two users each achieving a target SIR $N_{\rm I}\gamma$ is equivalent to the FAMA network with $N_{\rm I}+1$ users, each achieving a target SIR $\gamma$. The following theorem illustrates the intriguing fact that FAMA prefers serving more users with a less SIR target than serving less users with a more stringent SIR target.

\begin{theorem}\label{th:morebetter}
Based on the outage capacity lower bound (\ref{eqn:C-fama}), FAMA delivers higher capacity by serving more users, with a less SIR target  because
\begin{equation}\label{eqn:morebetter}
\frac{\left.C_{\rm FAMA}(\gamma)\right|_{N_{\rm I}+1}}{\left.C_{\rm FAMA}(N_{\rm I}\gamma)\right|_{2}}\approx\left(\frac{N_{\rm I}+1}{2}\right)\frac{1}{1+\log_{\gamma}N_{\rm I}},
\end{equation}
where $\left.C_{\rm FAMA}(\gamma)\right|_{M}$ denotes the outage capacity lower bound for the FAMA network with $M$ simultaneous users each achieving a target SIR of $\gamma$.
\end{theorem}

\proof As $\varepsilon_{\rm UB}(\gamma)$ is the same for the two cases, the ratio immediately gives (\ref{eqn:morebetter}) if $\gamma$ is large.
\endproof

For example, if $\gamma=10$ and $N_{\rm I}=100$, $\frac{\left.C_{\rm FAMA}(\gamma)\right|_{N_{\rm I}+1}}{\left.C_{\rm FAMA}(N_{\rm I}\gamma)\right|_{2}}\approx 17$, meaning that we can have $17$ times more capacity for serving $101$ users each with a target SIR of $10$ than two users with a target SIR of $1000$. The gain is not surprising because SIR is inside the log of the capacity formula while the number of users scales the capacity directly. What is less intuitive is how the excessive inter-user interference gets resolved when $N_{\rm I}$ is large to sustain the overall capacity rise. Under the notion of utilizing deep fades for multiple access, it is actually desirable to have superposition of a larger number of user signals, as this can be translated into deeper fades on the signal envelope and more opportunity for multiple access. Nevertheless, if the number of interferers increases, it usually comes with an increase in the total interference power which is indeed undesirable. These phenomena will be further characterized by the following corollary.

\begin{Corollary}\label{co:mu}
For a sufficiently large but fixed $N$ at every FAMA user, FAMA achieves the multiplexing gain $m$ or more with a target SIR $\gamma$ if the autocorrelation parameter at the FAMA user satisfies
\begin{equation}\label{eqn:mucon}
\mu\le\mu^*=\sqrt{\frac{\sum_{k=1}^{N-1}\frac{{N-1 \choose k}(-1)^{k+1}}{\left(\frac{\sigma_{\rm I}^2\gamma}{\sigma^2}\right)^k}-\frac{m}{N_{\rm I}+1}}{\sum_{k=1}^{N-1}\frac{k{N-1 \choose k}(-1)^{k+1}}{\left(\frac{\sigma_{\rm I}^2\gamma}{\sigma^2}\right)^k}}}\approx\sqrt{1-\frac{m\left(\frac{\sigma_{\rm I}^2\gamma}{\sigma^2}\right)}{(N_{\rm I}+1)(N-1)}}.
\end{equation}
\end{Corollary}

\proof This comes directly from (\ref{eqn:Ncon-2}) and (\ref{eqn:Ncon-3}).
\endproof

A few interesting observations can be made from (\ref{eqn:mucon}). First of all, more demanding targets $\gamma$ and $m$ would require $\mu$ to be smaller, as expected. Also, the increasing number of ports $N$ has the impact of lessening the requirement on the autocorrelation. On the other hand, very interestingly, if the number of interferers $N_{\rm I}$ increases while keeping the total power $\sigma_{\rm I}^2$ fixed, it has a desirable effect on relaxing the requirement on $\mu$. This can be explained by the fact that $N_{\rm I}$ contributes to more multipath which in turn gives rise to more fluctuations on the signal envelope that can be exploited by FAMA.

Thus far, some physical insight has been gained but under the assumption that all the autocorrelation parameters are the same, i.e., $|\mu_2|=\cdots=|\mu_N|=\mu$, which unfortunately may not correspond to any real configuration. The next theorems address the general case, and link the dimension of the fluid antenna, $W\lambda$, at a FAMA user to the other important network parameters.

\begin{theorem}\label{th:W}
The minimum required size of the fluid antenna at each FAMA user to achieve the targets $m$ and $\gamma$ for a given $N_{\rm I}$ and large but fixed $N$ is given by
\begin{equation}\label{eqn:Wmin}
W\ge \frac{1}{\pi}J_0^{-1}\left(\sqrt{1-\frac{m\left(\frac{\sigma_{\rm I}^2\gamma}{\sigma^2}\right)}{(N_{\rm I}+1)\left(\left\lfloor\frac{N}{2}\right\rfloor-1\right)}}\right),
\end{equation}
where $J_0^{-1}(\cdot)$ computes the inverse of $J_0(\cdot)$ and is defined in such a way that $\rho^*=J_0^{-1}(\mu^*)$ returns the minimum value of $\rho^*$ to ensure that $|J_0(\rho)|\le\mu^*$ for $\rho\ge\rho^*$. Additionally, the notation $\lfloor\cdot\rfloor$ returns the largest integer that is smaller than the input.
\end{theorem}

\proof See Appendix E.
\endproof

\begin{theorem}\label{th:Ngeneral}
For the general case, the number of ports $N$ for each FAMA user should satisfy
\begin{equation}\label{eqn:Ngeneral}
N\ge2\left[\frac{m\left(\frac{\sigma_{\rm I}^2\gamma}{\sigma^2}\right)}{(N_{\rm I}+1)\left(1-J_0^2(\pi W)\right)}+1\right].
\end{equation}
\end{theorem}

\proof This result is obtained by changing the subject of the condition (\ref{eqn:Wmin}) as $N$.
\endproof

From (\ref{eqn:Ngeneral}), we can see that the power of FAMA comes predominately from the number of ports, $N$, though $W$ has an impact on the required value of $N$ to meet the SIR target $\gamma$ and the multiplexing gain $m$. The result in Theorem \ref{th:Ngeneral} suggests that theoretically, as long as $W\ne 0$, there always exists a finite value of $N$ such that FAMA achieves any arbitrarily given multiplexing gain $m$ with any given target SIR $\gamma$. On a negative note, however, the required value of $N$ tends to be quite large. For example, assuming $\sigma_{\rm I}^2=N_{\rm I}\sigma^2$, even if $W$ is so large that $J_0(\pi W)\approx 0$, according to (\ref{eqn:Ngeneral}), we still need
\begin{equation}
N\approx 2m\gamma+2=2(2)(10)+2=42~\mbox{if }m=2~\mbox{and }\gamma=10.
\end{equation}
The above estimate is, however, based on a sufficient condition and the outage probability upper bound. Therefore, the actual minimum required $N$ may be smaller. 

\begin{Corollary}\label{co:m-FAMA}
The multiplexing gain of the FAMA network for the general case is given by
\begin{equation}\label{eqn:m-FAMA-gen}
m\approx\min\left\{\frac{\left(\frac{N}{2}-1\right)\left(1-J_0^2(\pi W)\right)(N_{\rm I}+1)}{\left(\frac{\sigma_{\rm I}^2\gamma}{\sigma^2}\right)},N_{\rm I}+1\right\}\approx\min\left\{\frac{\left(\frac{N}{2}-1\right)\left(1-J_0^2(\pi W)\right)}{\gamma},N_{\rm I}+1\right\}.
\end{equation}
\end{Corollary}

\proof The result is obtained from (\ref{eqn:m-FAMA}) by using the same argument used in Appendix E.
\endproof

Corollary \ref{co:m-FAMA} illustrates the potential capability of FAMA whose average outage capacity scales linearly with the number of ports, $N$, but is discounted by the SIR target $\gamma$. The size of the fluid antenna, $W$, contributes to the amount of spatial correlation that has a critical impact on $m$ only if $W$ is very small. For example, if $W=0.5$, $J_0^2(0.5\pi)\approx 0.22$ which is already small, and will not be a major factor for the capacity performance. This reveals that space being the major limitation due to spatial correlation may be overstated, and massive capacity increase is possible if $N$ can grow to be large. Note, however, that according to the definition (\ref{eqn:mg}), $m$ will always be limited by the number of users $N_{\rm I}+1$.


\section{Numerical Results}
In this section, we provide some numerical results for discussion. As there are many system parameters involved in the simulations, many results are presented but our aim is to highlight some key observations and interested readers are welcome to examine the results more closely. The results are meant to provide insight on the possible performance of FAMA from an information-theoretic perspective and may not represent an accurate evaluation of FAMA in practice given the assumptions used in our model. In the simulations, it is assumed that the users are statistically identical and that $\sigma_{\rm I}^2=N_{\rm I}\sigma^2$.

\begin{figure}[]
\begin{center}
\subfigure[$W=0.5$]{\includegraphics[width=12cm]{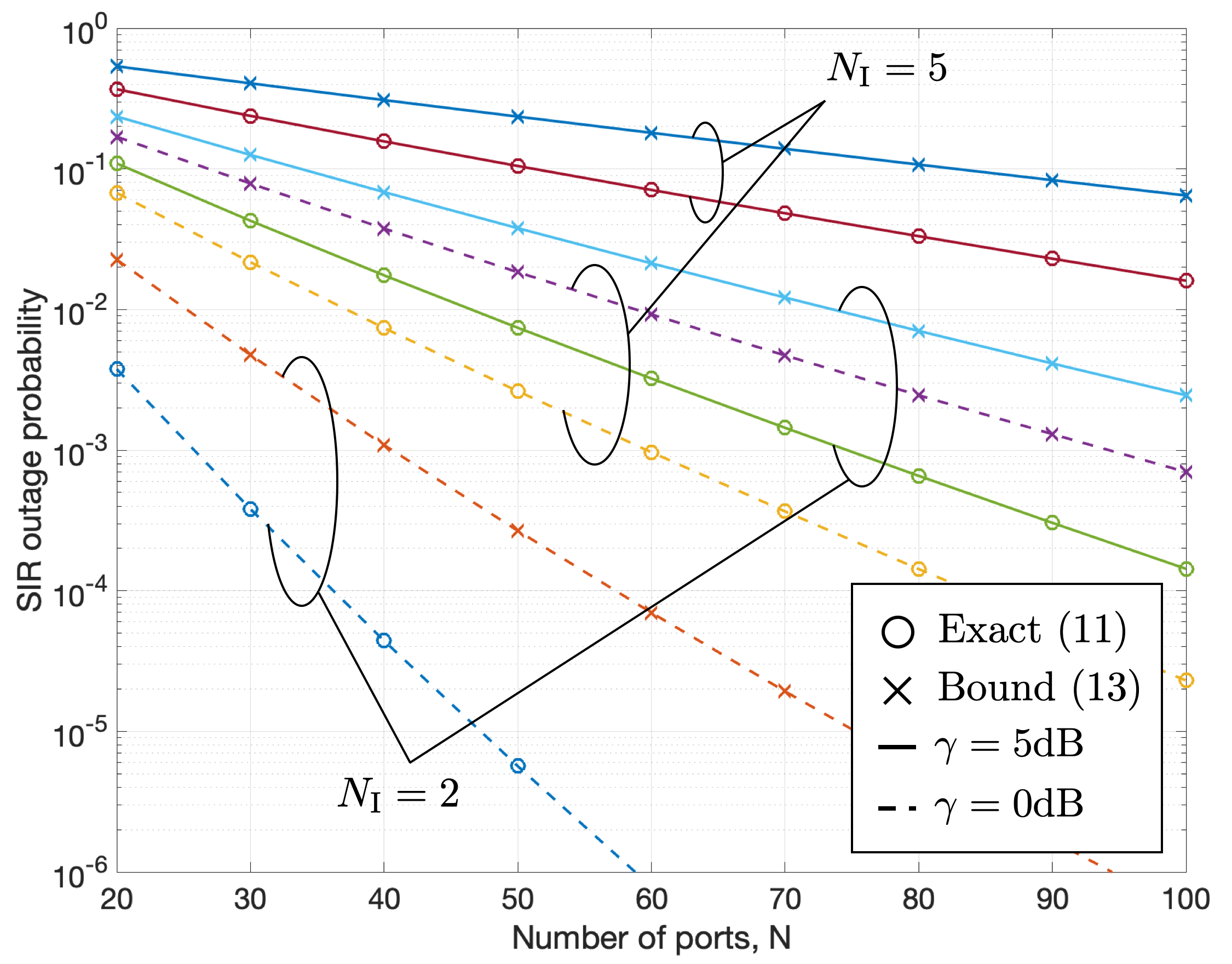}\label{fig:fama-exVSub-a}}
\subfigure[$W=2$]{\includegraphics[width=12cm]{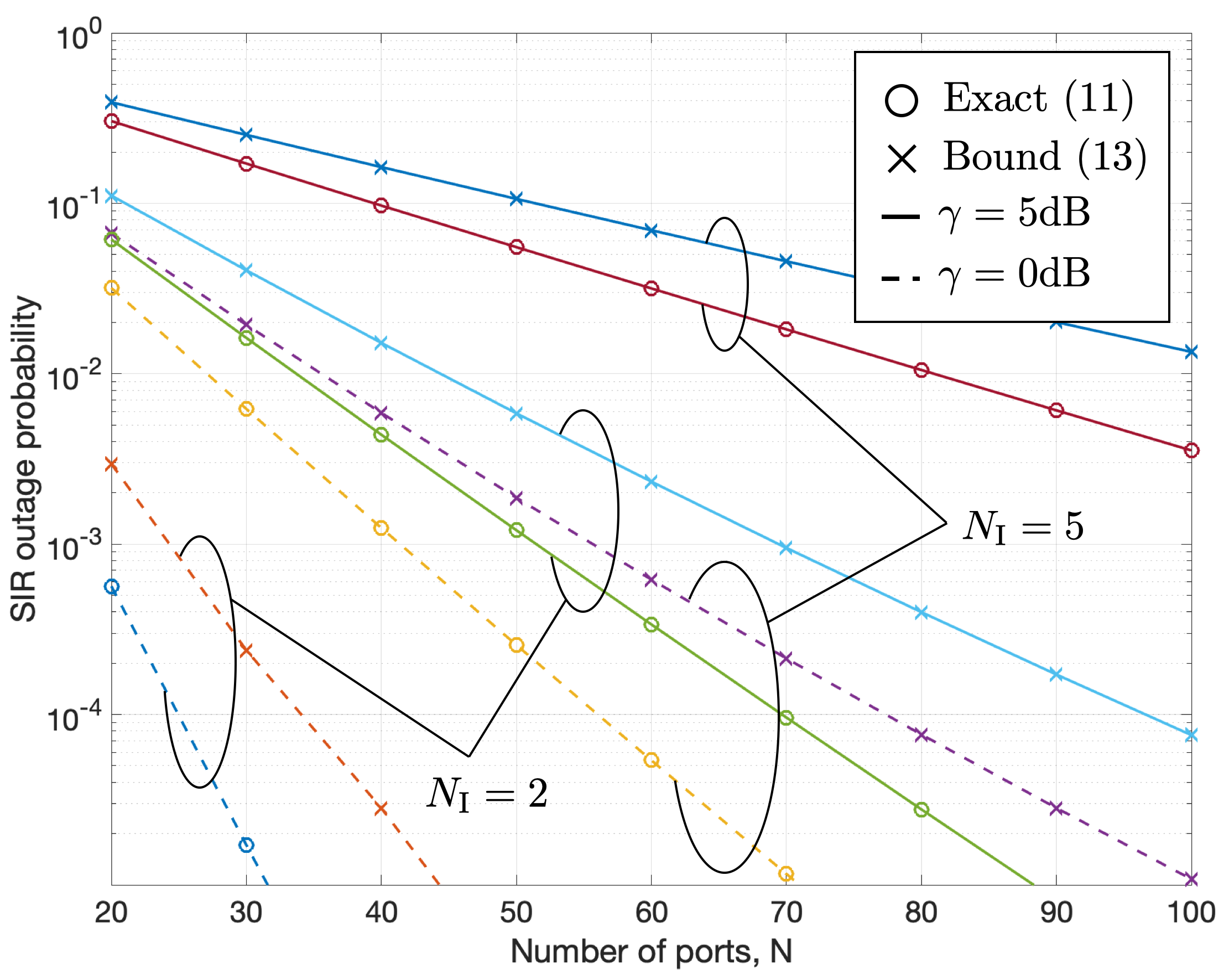}\label{fig:fama-exVSub-b}}
\caption{Comparison of the exact SIR outage probability (\ref{eqn:op-sir}) and the upper bound (\ref{eqn:op-ub1}).}\label{fig:fama-exVSub}
\end{center}
\end{figure}

Results in Figure \ref{fig:fama-exVSub} compare the exact SIR outage probability in (\ref{eqn:op-sir}) and the upper bound in (\ref{eqn:op-ub1}) for various different settings. It can be seen that the upper bound is not particularly tight but can be interpreted as a conservative measure to the achievable performance of FAMA and used as a means to understand analytically the interplay between different system parameters. In particular, the upper bound will be tighter if the size of the fluid antenna $W$ is larger, or the operating conditions are challenging, e.g., with a larger number of interferers $N_{\rm I}$ and/or a large SIR target $\gamma$. It should be noted that in terms of the network outage capacity, we tend to focus on larger values of outage probability (e.g., $\approx 1\%-10\%$)\footnote{This is because the impact on the average outage capacity will be negligible if the outage probability is smaller than $1\%$.} where the bound appears to pick up the slope of the outage probability quite accurately, although the bound always overestimates the required number of ports by a margin. The results in Figure \ref{fig:fama-exVSub-b} show some encouraging results even if $N$ is not so large. It is observed that with $W=2$, $N=20$ and $N_{\rm I}=5$ (i.e., supporting $6$ users), the SIR outage probability at a FAMA user is about $30\%$, which means that FAMA achieves a multiplexing gain of $6\times(1-0.3)=4.2$. If $N=30$, the multiplexing gain is increased to $4.8$.

\begin{figure}[]
\begin{center}
\includegraphics[width=12cm]{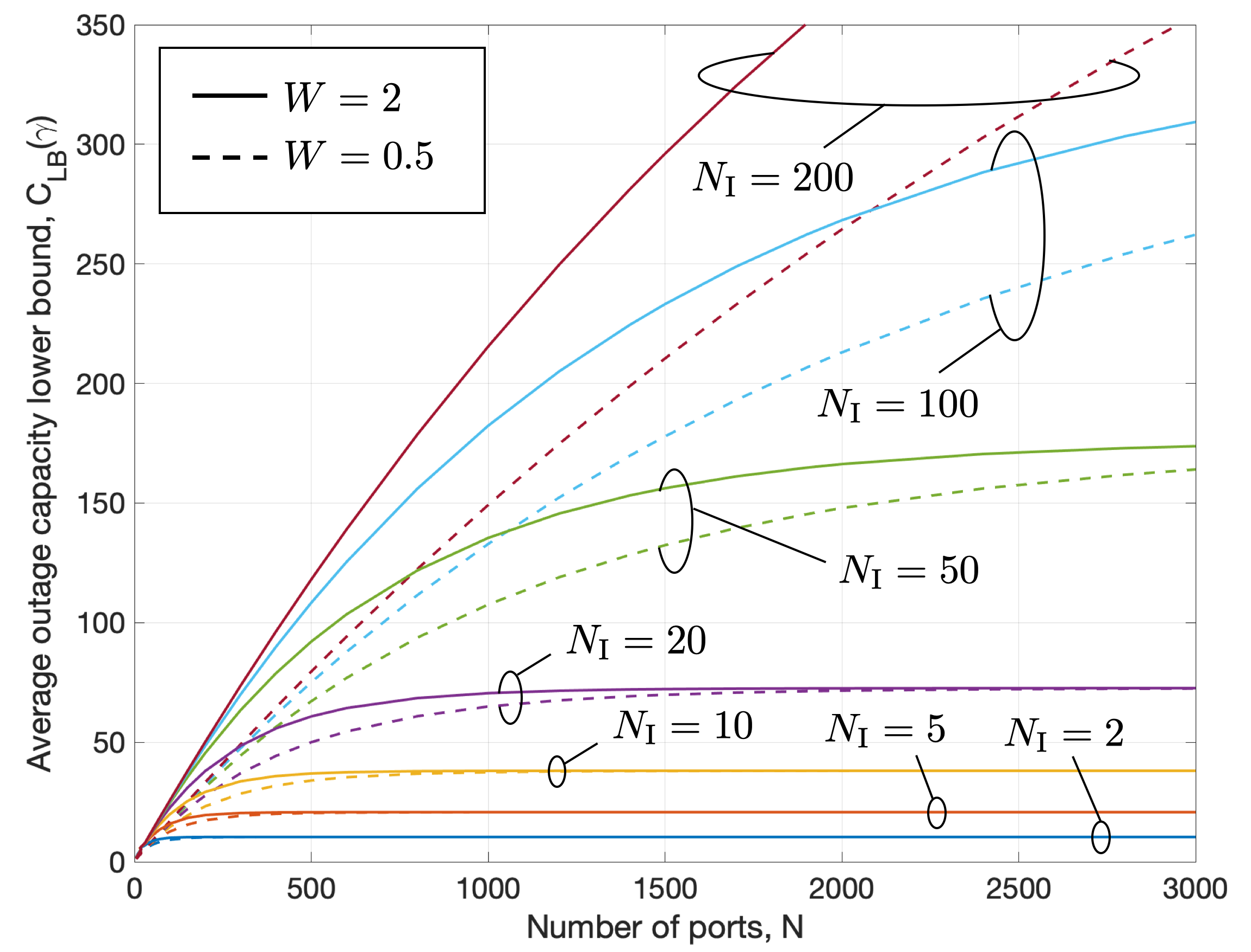}
\caption{The network outage capacity lower bound in (\ref{eqn:C-fama}) versus $N$ when $\gamma=10{\rm dB}$.}\label{fig:fama-CvsN}
\end{center}
\end{figure}

From now on, we illustrate only the numerical results based on the upper bound (\ref{eqn:op-ub1}) since the complexity to evaluate (\ref{eqn:op-sir}) can quickly become unmanageable if $N$ is large. Although this will only provide a conservative view on the performance of FAMA, the study of the average outage capacity performance and multiplexing gain would be reasonably accurate as only large values of outage probability matter. Results in Figure \ref{fig:fama-CvsN} are provided for the average network outage capacity of FAMA when each user has an SIR target of $10{\rm dB}$. As we can see, a larger $W$ will achieve higher capacity but the difference gradually disappears if $N$ grows to be extremely large. Increasing $N$ apparently enhances capacity but the gain will saturate at some point which depends on the total number of users, $N_{\rm I}+1$. In particular, the maximum average outage capacity is $(N_{\rm I}+1)\log_2(1+\gamma)$ which is achievable when $N\to\infty$ (so achieving zero outage probability). As a result, as expected, we observe that if $N_{\rm I}$ increases, FAMA delivers greater capacity, while the interference at each user can be handled by a fluid antenna with sufficiently large $N$.

\begin{figure}[]
\begin{center}
\includegraphics[width=12cm]{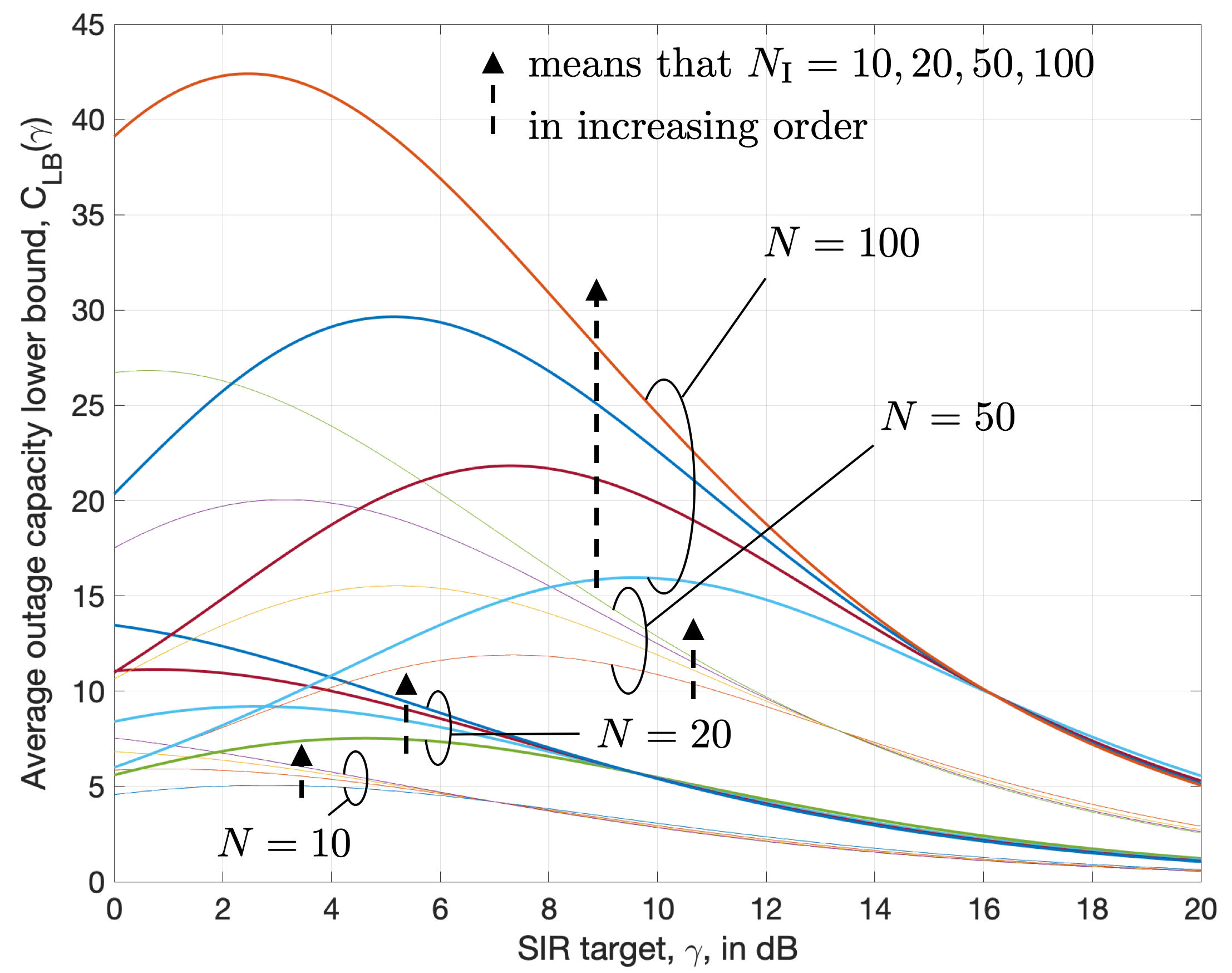}
\caption{The network outage capacity lower bound in (\ref{eqn:C-fama}) versus the SIR target $\gamma$ when $W=2$.}\label{fig:fama-CvsG}
\end{center}
\end{figure}

We investigate the impact of the target SIR $\gamma$ using the results in Figure \ref{fig:fama-CvsG}. The results demonstrate that for large $N$, there is an optimal $\gamma$ that maximizes the average outage capacity of the network. A too small $\gamma$ unnecessarily limits the achievable performance for each user but if $\gamma$ is too large, the outage probability will begin to increase drastically and reduce the overall network outage capacity. The number of interferers $N_{\rm I}$ generally has a positive impact on the outage capacity as discussed above but if $\gamma$ is too large for a given $N$, too many interferers will harm the overall capacity. Such turning point appears to happen at a larger value of $\gamma$ if $N$ is greater. Despite this, after the turning point, the network outage capacity is not much affected by the number of interferers. There are also some highlights in this figure. For $\gamma=0{\rm dB}$ and $N=20$, a FAMA network with $101$ users can deliver a multiplexing gain of $13$ (note that $\log_2(1+\gamma)=1$ in this case) which will be doubled to $26$ as a result of reduction in the SIR outage probability if $N$ is increased to $50$. Thus, a massive capacity gain is possible for not so large $N$.

\begin{figure}[]
\begin{center}
\includegraphics[width=12cm]{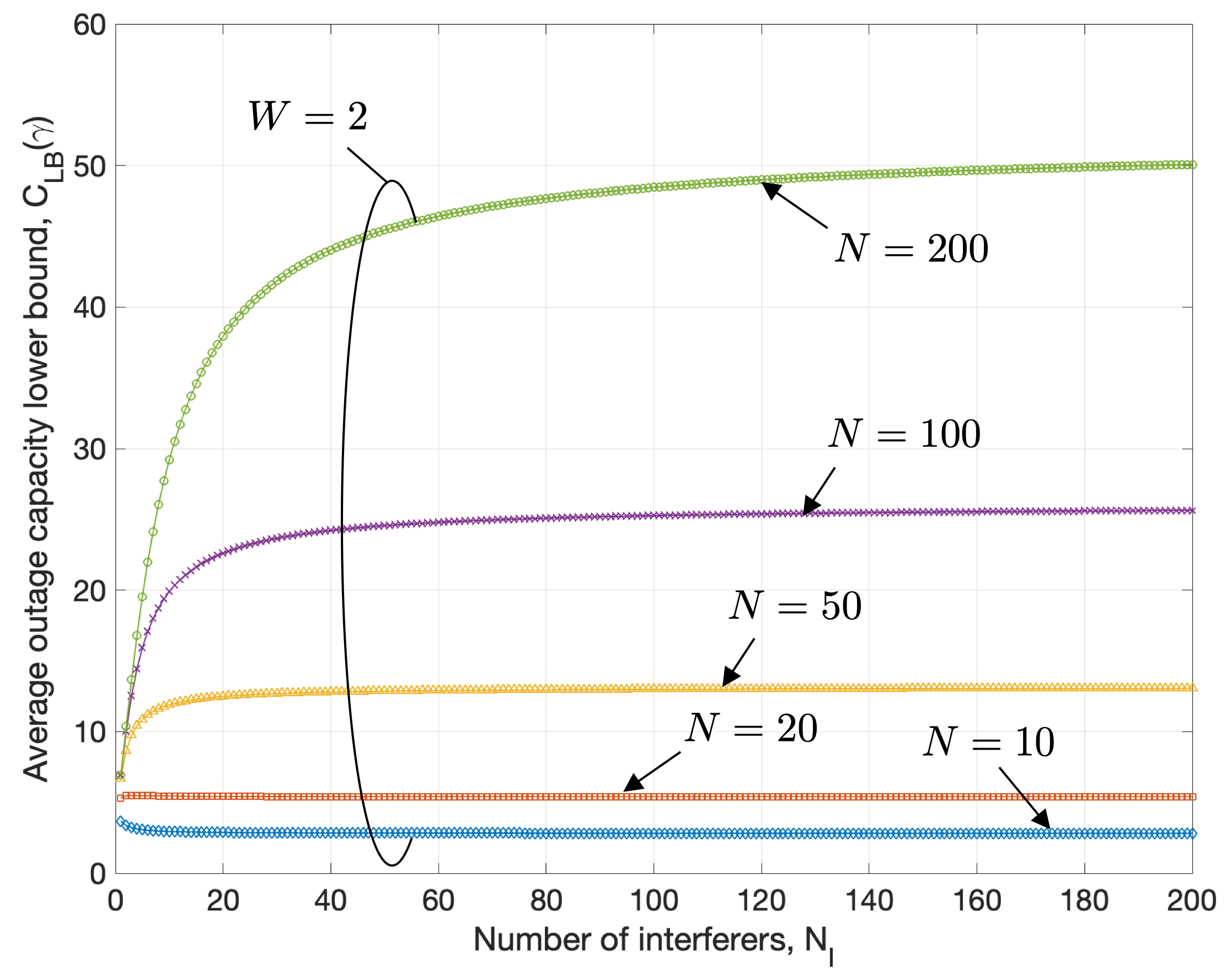}
\caption{The outage capacity lower bound in (\ref{eqn:C-fama}) versus $N_{\rm I}$ when $\gamma=10{\rm dB}$.}\label{fig:fama-CvsNI}
\end{center}
\end{figure}

The impact of $N_{\rm I}$ is further examined by the results in Figure \ref{fig:fama-CvsNI} where $\gamma=10{\rm dB}$ is considered. As can be seen, when $N$ is too small, the outage capacity first decreases and then remains the same as $N_{\rm I}$ increases. For larger values of $N$, in contrast, the outage capacity grows as $N_{\rm I}$ increases but saturates at some point because the outage capacity gain by the increase in the number of users is cancelled by the corresponding increase in the outage probability at each user when $N$ is fixed. 

\begin{figure}[]
\begin{center}
\includegraphics[width=12cm]{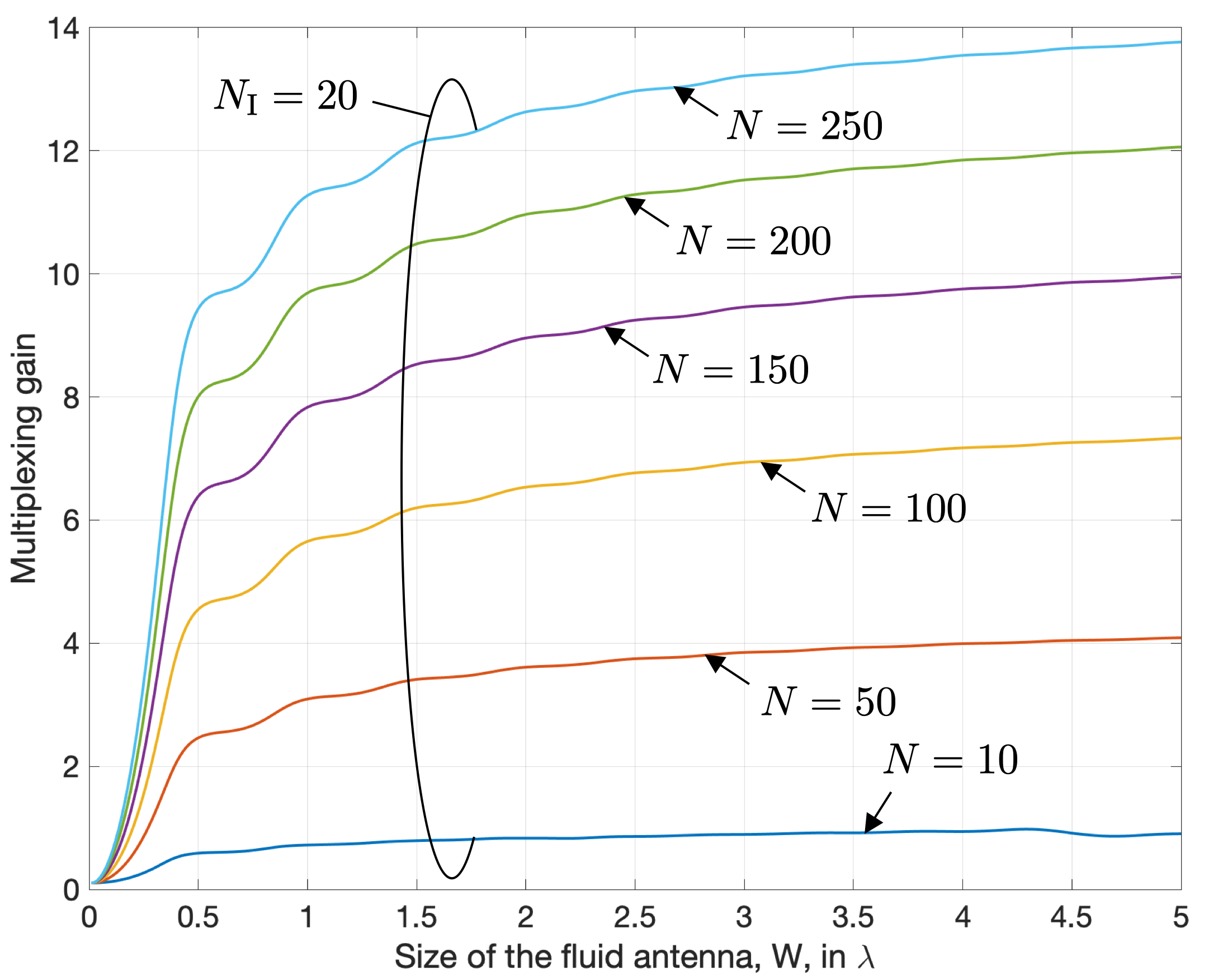}
\caption{The multiplexing gain in (\ref{eqn:mg}) versus the size of the fluid antenna $W$ when $\gamma=10{\rm dB}$.}\label{fig:ffama-mvsW}
\end{center}
\end{figure}

Results in Figure \ref{fig:ffama-mvsW} study the multiplexing gain of the FAMA network against the size of the fluid antenna at each user.  The results illustrate that the impact of $W$ is more obvious when $N$ is larger. As expected, if $W$ increases, the multiplexing gain will follow. The capacity benefit is most sharp before $\frac{\lambda}{2}$ but after $\frac{\lambda}{2}$, the increase in the multiplexing gain diminishes. Putting implementation issues aside, this suggests that $\frac{\lambda}{2}$ be the smallest size that can obtain the most benefit of fluid antenna. Moreover, some ripples are observed on the shape of the curves and these come naturally from how the autocorrelation varies over a linear distance which follows a Bessel function. Also, the results again indicate that as $N$ increases, a greater multiplexing gain is achieved, and clearly $N$ is a more important factor than $W$.

\begin{figure}[]
\begin{center}
\includegraphics[width=12cm]{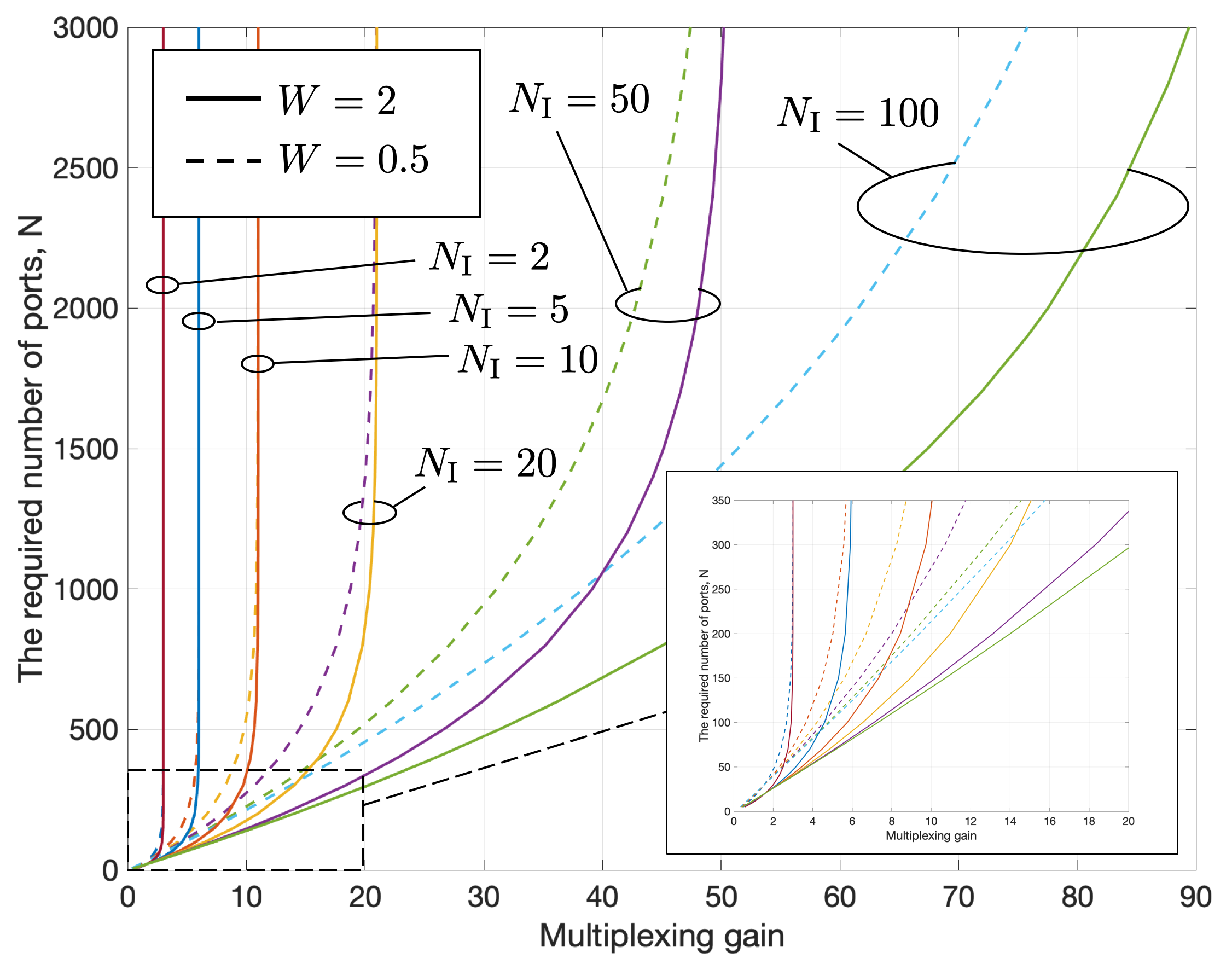}
\caption{The required number of ports, $N$, versus the multiplexing gain in (\ref{eqn:mg}) when $\gamma=10{\rm dB}$.}\label{fig:fama-Nvsm}
\end{center}
\end{figure}

Multiplexing gain measures the capacity increase in the network. Results in Figure \ref{fig:fama-Nvsm} investigate the required number of ports against the achievable multiplexing gain. First of all, it should be noted that the multiplexing gain is upper bounded by $N_{\rm I}+1$ (the number of users) which is achievable only when the outage probability goes to zero if $N\to\infty$. The same can be observed from the results in this figure, as $N$ continues to increase while keeping $N_{\rm I}$ fixed. The results also indicate that the benefit of a larger $W$ gets smaller as $N$ can be increased to compensate for the loss of outage probability performance. Another interesting observation is that if $N$ can be extremely large, then FAMA can accommodate hundreds of users, all by a single fluid antenna with a reasonable size at each user, and obtain a massive multiplexing gain; see also Figure \ref{fig:fama-CvsN}. Furthermore, we can see the results for more practical values of $N$ from the magnified figure, which reveals that as large as a multiplexing gain of $10$ is possible when $N=150$. Note that the results are based on the outage probability upper bound which will overestimate the value of $N$ so it is anticipated that a smaller $N$ will likely achieve the impressive multiplexing gain.

\begin{figure}[]
\begin{center}
\includegraphics[width=12cm]{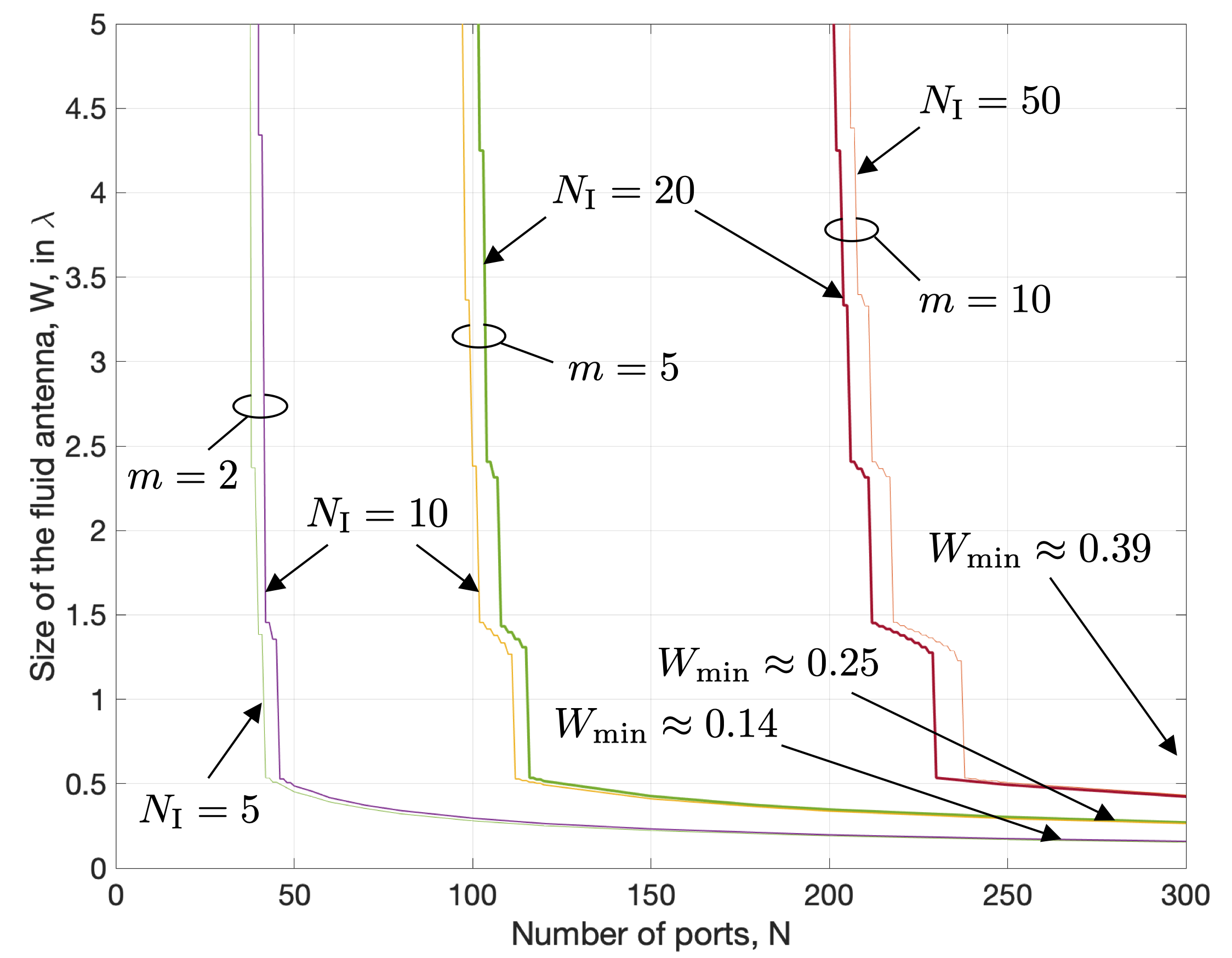}
\caption{The required size of fluid antenna, $W$, in (\ref{eqn:Wmin}) versus the number of ports when $\gamma=10{\rm dB}$.}\label{fig:fama-WvsN}
\end{center}
\end{figure}

We now take a closer look at the required size of the fluid antenna for a FAMA user as a function of $N$ to achieve a certain multiplexing gain of the network. The results are shown in Figure \ref{fig:fama-WvsN}, which indicates that if $N$ is too small, then it is impossible to have a feasible size $W$ that can achieve a given multiplexing gain. Nonetheless, once $N$ is sufficiently large, $W$ goes down sharply and then gradually decreases as $N$ continues to increase. Results also illustrate that the number of interferers does not have a significant impact on $W$ and $W$ is mainly dependent on the required multiplexing gain. In addition, $N$ needs an exponential increase in values to reduce the required size of the fluid antenna beyond $\frac{\lambda}{2}$.

\section{Conclusion}
This paper proposed a novel concept for multiple access, referred to as FAMA, which handles inter-user interference purely by scanning through the fading envelopes and picking the best over a number of ports closely located within a small linear space. FAMA is inspired by the intuition that all signals including the interference suffer from deep fades, and multiple access is possible if the moments of deep fades (in space) for interference are exploited. The concept is also motivated by the emerging fluid antenna technology for software-controlled position-flexible antennas. Theoretically, it was shown that if the number of ports is sufficiently large, a fluid antenna can achieve any arbitrarily small SIR outage probability, illustrating its feasibility for interference elimination. Also, we analyzed the average outage capacity and the multiplexing gain of the FAMA network. In particular, we showed that the network multiplexing gain grows linearly with the number of ports at each user while it is ultimately limited by the number of users. Our results revealed that the size of the fluid antenna has most impact if it is smaller than $\frac{\lambda}{2}$ but its impact has a diminishing return in the network capacity if it grows beyond $\frac{\lambda}{2}$. Additionally, our results demonstrated that accommodating hundreds of users in the same radio resource is possible, all by a single fluid-antenna with a small size at each user in FAMA, thereby achieving a significant capacity gain.

\section*{Appendices}
\subsection{Proof of Theorem \ref{th:op-fama}}
By definition, we have
\begin{equation}
{\rm Prob}({\rm SIR}\le\gamma)={\rm Prob}\left(\max_k\left\{\frac{|g_k|^2}{|g_k^{\rm I}|^2}\right\}\le\gamma\right)={\rm Prob}\left(\frac{|g_1|}{|g_1^{\rm I}|}\le\sqrt{\gamma},\frac{|g_2|}{|g_2^{\rm I}|}\le\sqrt{\gamma},\dots,\frac{|g_N|}{|g_N^{\rm I}|}\le\sqrt{\gamma}\right),
\end{equation}
which further gives
\begin{multline}
{\rm Prob}({\rm SIR}\le\gamma)=\int_0^\infty\dots\int_0^\infty{\rm Prob}\left(\left.|g_1|\le\sqrt{\gamma}t_1,\dots,|g_N|\le\sqrt{\gamma}t_N\right|t_1,\dots,t_N\right)\times\\
p_{|g_1^{\rm I}|,\dots,|g_N^{\rm I}|}(t_1,\dots,t_N)dt_1\cdots dt_N.
\end{multline}
The conditional cumulative distribution function (cdf), ${\rm Prob}\left(\left.|g_1|\le\sqrt{\gamma}t_1,\dots,|g_N|\le\sqrt{\gamma}t_N\right|t_1,\dots,t_N\right)$, can be obtained by \cite[Theorem 2]{Wong-2020twc} while the pdf, $p_{|g_1^{\rm I}|,\dots,|g_N^{\rm I}|}(t_1,\dots,t_N)$, is given by \cite[Theorem 1]{Wong-2020twc}. As a consequence, we obtain
\begin{multline}\label{eqn:a-0}
{\rm Prob}({\rm SIR}\le\gamma)=\int_{t_1=0}^\infty\frac{2t_1}{\sigma_{\rm I}^2}e^{-\frac{t_1^2}{\sigma_{\rm I}^2}}\int_{t=0}^\frac{\gamma t_1^2}{\sigma^2} e^{-t}\prod_{k=2}^N\Bigg\{\\
\left.\int_{t_k=0}^\infty
\left[1-Q_1\left(\sqrt{\frac{2\mu_k^2}{1-\mu_k^2}}\sqrt{t},\sqrt{\frac{2}{\sigma^2(1-\mu_k^2)}}\sqrt{\gamma}t_k\right)\right]
\frac{2t_k}{\sigma_{\rm I}^2(1-\mu_k^2)}e^{-\frac{t_k^2+\mu_k^2t_1^2}{\sigma_{\rm I}^2(1-\mu_k^2)}}
I_0\left(\frac{2\mu_k t_1 t_k}{\sigma_{\rm I}^2(1-\mu_k^2)}\right)dt_k\right\} dt dt_1.
\end{multline}
To evaluate the integration over $t_k$, we first note that
\begin{equation}\label{eqn:a-1}
\int_{t_k=0}^\infty
\frac{2t_k}{\sigma_{\rm I}^2(1-\mu_k^2)}e^{-\frac{t_k^2+\mu_k^2t_1^2}{\sigma_{\rm I}^2(1-\mu_k^2)}}
I_0\left(\frac{2\mu_k t_1 t_k}{\sigma_{\rm I}^2(1-\mu_k^2)}\right)dt_k=1
\end{equation}
because this is the total probability for a Rician random variable. Then the following lemma is useful.

\begin{lemma}\label{lemma:B32}
The following identity is true:
\begin{equation}
\int_0^\infty x e^{-\frac{x^2}{2}}I_0(cx)Q_1(b,ax)dx=e^\frac{c^2}{2}Q_1\left(\frac{b}{\sqrt{a^2+1}},\frac{ac}{\sqrt{a^2+1}}\right)-\frac{a^2}{a^2+1}e^\frac{c^2-b^2}{2(a^2+1)}I_0\left(\frac{abc}{a^2+1}\right).
\end{equation}
\end{lemma}

\proof This is the result in \cite[(B.32)]{Simon-2002} when $p=1$.
\endproof

Considering the other term under the integration of $t_k$ and changing the variable by $x=\frac{t_k}{\sqrt{\frac{\sigma_{\rm I}^2}{2}(1-\mu_k^2)}}$, we get
\begin{multline}
\int_{0}^\infty
Q_1\left(\sqrt{\frac{2\mu_k^2}{1-\mu_k^2}}\sqrt{t},\sqrt{\frac{2}{\sigma^2(1-\mu_k^2)}}\sqrt{\gamma}t_k\right)
\frac{2t_k}{\sigma_{\rm I}^2(1-\mu_k^2)}e^{-\frac{t_k^2+\mu_k^2t_1^2}{\sigma_{\rm I}^2(1-\mu_k^2)}}
I_0\left(\frac{2\mu_k t_1 t_k}{\sigma_{\rm I}^2(1-\mu_k^2)}\right)dt_k\\
=e^{-\frac{\mu_k^2 t_1^2}{\sigma_{\rm I}^2(1-\mu_k^2)}}\int_0^\infty x e^{-\frac{x^2}{2}}I_0\left(\frac{\mu_kt_1}{\sqrt{\frac{\sigma_{\rm I}^2}{2}(1-\mu_k^2)}}x\right)Q_1\left(\sqrt{\frac{2\mu_k^2}{1-\mu_k^2}}\sqrt{t},\frac{\sigma_{\rm I}\sqrt{\gamma}}{\sigma}x\right)dx.
\end{multline}
Now, using the result in Lemma \ref{lemma:B32} on the right hand side, it yields
\begin{multline}\label{eqn:a-2}
\int_{0}^\infty
Q_1\left(\sqrt{\frac{2\mu_k^2}{1-\mu_k^2}}\sqrt{t},\sqrt{\frac{2}{\sigma^2(1-\mu_k^2)}}\sqrt{\gamma}t_k\right)
\frac{2t_k}{\sigma_{\rm I}^2(1-\mu_k^2)}e^{-\frac{t_k^2+\mu_k^2t_1^2}{\sigma_{\rm I}^2(1-\mu_k^2)}}
I_0\left(\frac{2\mu_k t_1 t_k}{\sigma_{\rm I}^2(1-\mu_k^2)}\right)dt_k\\
=Q_1\left(\frac{1}{\sqrt{\frac{\sigma_{\rm I}^2\gamma}{\sigma^2}+1}}\sqrt{\frac{2\mu_k^2}{1-\mu_k^2}}\sqrt{t},\sqrt{\frac{\frac{\sigma_{\rm I}^2\gamma}{\sigma^2}}{\frac{\sigma_{\rm I}^2\gamma}{\sigma^2}+1}}\sqrt{\frac{2\mu_k^2}{1-\mu_k^2}}t_1\right)\\
-\left(\frac{\frac{\sigma_{\rm I}^2\gamma}{\sigma^2}}{\frac{\sigma_{\rm I}^2\gamma}{\sigma^2}+1}\right)e^{-\left(\frac{1}{\frac{\sigma_{\rm I}^2\gamma}{\sigma^2}+1}\right)\frac{\mu_k^2}{1-\mu_k^2}\left(\frac{\gamma t_1^2}{\sigma^2}+t\right)}
I_0\left(\frac{\frac{\sigma_{\rm I}\sqrt{\gamma}}{\sigma}}{\frac{\sigma_{\rm I}^2\gamma}{\sigma^2}+1}\left(\frac{2\mu_k^2}{\sigma_{\rm I}(1-\mu_k^2)}\right)t_1\sqrt{t}\right).
\end{multline}
Using (\ref{eqn:a-1}) and (\ref{eqn:a-2}) into (\ref{eqn:a-0}), and changing the variable by $z=\frac{t_1^2}{\sigma_{\rm I}^2}$ give the final result.

\subsection{Proof of Theorem \ref{th:op-ub1}}
To start with, we need the following lemma.

\begin{lemma}\label{lemma:Q1lb}
We have the following lower bound for $Q_1(\alpha,\beta)$:
\begin{equation}
Q_1(\alpha,\beta)\ge e^{-\frac{\alpha^2+\beta^2}{2}}I_0(\alpha\beta).
\end{equation}
\end{lemma}

\proof This can be obtained by taking only the first term of the definition in \cite[(A.5)]{Simon-2002}.
\endproof

Using Lemma \ref{lemma:Q1lb} on the $Q_1(\cdot,\cdot)$ term inside the integration of (\ref{eqn:op-sir}) gives
\begin{multline}
Q_1\left(\frac{1}{\sqrt{\frac{\sigma_{\rm I}^2\gamma}{\sigma^2}+1}}\sqrt{\frac{2\mu_k^2}{1-\mu_k^2}}\sqrt{t},\sqrt{\frac{\frac{\sigma_{\rm I}^2\gamma}{\sigma^2}}{\frac{\sigma_{\rm I}^2\gamma}{\sigma^2}+1}}\sqrt{\frac{2\mu_k^2}{1-\mu_k^2}}\sqrt{z}\right)\\
\ge e^{-\left(\frac{1}{\frac{\sigma_{\rm I}^2\gamma}{\sigma^2}+1}\right)\frac{\mu_k^2}{1-\mu_k^2}\left(\frac{\sigma_{\rm I}^2\gamma}{\sigma^2}z+t\right)}I_0\left(\frac{\frac{\sigma_{\rm I}\sqrt{\gamma}}{\sigma}}{\frac{\sigma_{\rm I}^2\gamma}{\sigma^2}+1}\left(\frac{2\mu_k^2}{1-\mu_k^2}\right)\sqrt{zt}\right).
\end{multline}
Then apply the above result and evaluate the difference 
\begin{align}
&\left(\frac{\frac{\sigma_{\rm I}^2\gamma}{\sigma^2}}{\frac{\sigma_{\rm I}^2\gamma}{\sigma^2}+1}\right)e^{-\left(\frac{1}{\frac{\sigma_{\rm I}^2\gamma}{\sigma^2}+1}\right)\frac{\mu_k^2}{1-\mu_k^2}\left(\frac{\sigma_{\rm I}^2\gamma}{\sigma^2}z+t\right)}
I_0\left(\frac{\frac{\sigma_{\rm I}\sqrt{\gamma}}{\sigma}}{\frac{\sigma_{\rm I}^2\gamma}{\sigma^2}+1}\left(\frac{2\mu_k^2}{1-\mu_k^2}\right)\sqrt{zt}\right)\notag\\
&\hspace{+5cm}-Q_1\left(\frac{1}{\sqrt{\frac{\sigma_{\rm I}^2\gamma}{\sigma^2}+1}}\sqrt{\frac{2\mu_k^2}{1-\mu_k^2}}\sqrt{t},\sqrt{\frac{\frac{\sigma_{\rm I}^2\gamma}{\sigma^2}}{\frac{\sigma_{\rm I}^2\gamma}{\sigma^2}+1}}\sqrt{\frac{2\mu_k^2}{1-\mu_k^2}}\sqrt{z}\right)\notag\\
&\hspace{+2cm}\stackrel{(a)}{\le} \left(\frac{\frac{\sigma_{\rm I}^2\gamma}{\sigma^2}}{\frac{\sigma_{\rm I}^2\gamma}{\sigma^2}+1}\right)e^{-\left(\frac{1}{\frac{\sigma_{\rm I}^2\gamma}{\sigma^2}+1}\right)\frac{\mu_k^2}{1-\mu_k^2}\left(\frac{\sigma_{\rm I}^2\gamma}{\sigma^2}z+t\right)}I_0\left(\frac{\frac{\sigma_{\rm I}\sqrt{\gamma}}{\sigma}}{\frac{\sigma_{\rm I}^2\gamma}{\sigma^2}+1}\left(\frac{2\mu_k^2}{1-\mu_k^2}\right)\sqrt{zt}\right)\notag\\
&\hspace{+5cm}-e^{-\left(\frac{1}{\frac{\sigma_{\rm I}^2\gamma}{\sigma^2}+1}\right)\frac{\mu_k^2}{1-\mu_k^2}\left(\frac{\sigma_{\rm I}^2\gamma}{\sigma^2}z+t\right)}I_0\left(\frac{\frac{\sigma_{\rm I}\sqrt{\gamma}}{\sigma}}{\frac{\sigma_{\rm I}^2\gamma}{\sigma^2}+1}\left(\frac{2\mu_k^2}{1-\mu_k^2}\right)\sqrt{zt}\right)\notag\\
&\hspace{+2cm}\stackrel{(b)}{=}\left(\frac{-1}{\frac{\sigma_{\rm I}^2\gamma}{\sigma^2}+1}\right)e^{-\left(\frac{1}{\frac{\sigma_{\rm I}^2\gamma}{\sigma^2}+1}\right)\frac{\mu_k^2}{1-\mu_k^2}\left(\frac{\sigma_{\rm I}^2\gamma}{\sigma^2}z+t\right)}
I_0\left(\frac{\frac{\sigma_{\rm I}\sqrt{\gamma}}{\sigma}}{\frac{\sigma_{\rm I}^2\gamma}{\sigma^2}+1}\left(\frac{2\mu_k^2}{1-\mu_k^2}\right)\sqrt{zt}\right)\notag\\
&\hspace{+2cm}\stackrel{(c)}{\le}\left(\frac{-1}{\frac{\sigma_{\rm I}^2\gamma}{\sigma^2}+1}\right)\frac{e^{-\left(\frac{1}{\frac{\sigma_{\rm I}^2\gamma}{\sigma^2}+1}\right)\frac{\mu_k^2}{1-\mu_k^2}\left(\frac{\sigma_{\rm I}\sqrt{\gamma}}{\sigma}\sqrt{z}-\sqrt{t}\right)^2}}{1+4\left(\frac{1}{\frac{\sigma_{\rm I}^2\gamma}{\sigma^2}+1}\right)\left(\frac{\mu_k^2}{1-\mu_k^2}\right)\frac{\sigma_{\rm I}\sqrt{\gamma}}{\sigma}\sqrt{zt}}\notag\\
&\hspace{+2cm}\stackrel{(d)}{\le}\left(\frac{-1}{\frac{\sigma_{\rm I}^2\gamma}{\sigma^2}+1}\right)\frac{e^{-\left(\frac{1}{\frac{\sigma_{\rm I}^2\gamma}{\sigma^2}+1}\right)\left(\frac{\mu_k^2}{1-\mu_k^2}\right)\frac{\sigma_{\rm I}^2\gamma}{\sigma^2}z}}{1+4\left(\frac{1}{\frac{\sigma_{\rm I}^2\gamma}{\sigma^2}+1}\right)\left(\frac{\mu_k^2}{1-\mu_k^2}\right)\frac{\sigma_{\rm I}^2\gamma}{\sigma^2}z},\label{eqn:b-up}
\end{align}
where $(a)$ is from Lemma \ref{lemma:Q1lb}, $(c)$ uses the lower bound $I_0(x)\ge\frac{e^x}{1+2x}$ in \cite[(3.1)]{Yang-2016}, and $(d)$ substitutes $t=0$ in the numerator of $(c)$ and $t=\frac{\sigma_{\rm I}^2\gamma}{\sigma^2}z$ in the denominator of $(c)$. Now, using (\ref{eqn:b-up}) in the integration over $t$ in the outage probability (\ref{eqn:op-sir}) and performing a simple integration gives the final upper bound.

\subsection{Proof of Theorem \ref{th:op-ub2}}
Using the condition $|\mu_2|=\cdots=|\mu_N|=\mu$ and noting that $1-e^{-\frac{\sigma_{\rm I}^2\gamma}{\sigma^2}z}\le 1$, we have
\begin{equation}
{\rm Prob}({\rm SIR}\le\gamma)\le\varepsilon_{\rm UB}^{\rm I}\le\varepsilon_{\rm UB}^{\rm II}=\int_0^\infty e^{-z}\left[1-\left(\frac{1}{\frac{\sigma_{\rm I}^2\gamma}{\sigma^2}+1}\right)\frac{e^{-\left(\frac{1}{\frac{\sigma_{\rm I}^2\gamma}{\sigma^2}+1}\right)\left(\frac{\mu^2}{1-\mu^2}\right)\frac{\sigma_{\rm I}^2\gamma}{\sigma^2}z}}{1+4\left(\frac{1}{\frac{\sigma_{\rm I}^2\gamma}{\sigma^2}+1}\right)\left(\frac{\mu^2}{1-\mu^2}\right)\frac{\sigma_{\rm I}^2\gamma}{\sigma^2}z}\right]^{N-1}dz.
\end{equation}
Then by applying binomial expansion, we obtain
\begin{align}
\varepsilon_{\rm UB}^{\rm II}&=\int_0^\infty e^{-z}\sum_{k=0}^{N-1}{N-1\choose k}\frac{(-1)^k}{\left(\frac{\sigma_{\rm I}^2\gamma}{\sigma^2}+1\right)^k}\left[\frac{e^{-\left(\frac{1}{\frac{\sigma_{\rm I}^2\gamma}{\sigma^2}+1}\right)\left(\frac{\mu^2}{1-\mu^2}\right)\frac{\sigma_{\rm I}^2\gamma}{\sigma^2}z}}{1+4\left(\frac{1}{\frac{\sigma_{\rm I}^2\gamma}{\sigma^2}+1}\right)\left(\frac{\mu^2}{1-\mu^2}\right)\frac{\sigma_{\rm I}^2\gamma}{\sigma^2}z}\right]^k dz\notag\\
&=\sum_{k=0}^{N-1}\frac{{N-1\choose k}(-1)^k}{\left(\frac{\sigma_{\rm I}^2\gamma}{\sigma^2}+1\right)^k}\int_0^\infty
\left\{\frac{e^{-\left[\left(\frac{1}{\frac{\sigma_{\rm I}^2\gamma}{\sigma^2}+1}\right)\left(\frac{\mu^2}{1-\mu^2}\right)\frac{\sigma_{\rm I}^2\gamma}{\sigma^2}+\frac{1}{k}\right]z}}{1+4\left(\frac{1}{\frac{\sigma_{\rm I}^2\gamma}{\sigma^2}+1}\right)\left(\frac{\mu^2}{1-\mu^2}\right)\frac{\sigma_{\rm I}^2\gamma}{\sigma^2}z}\right\}^k dz.\label{eqn:c-ub}
\end{align}
Using the fact that $\int_0^\infty\left(\frac{e^{-ax}}{1+bx}\right)^k dx=\frac{e^\frac{ka}{b}}{b}E_k\left(\frac{ka}{b}\right)$ in (\ref{eqn:c-ub}) and assuming $\frac{\sigma_{\rm I}^2\gamma}{\sigma^2}\gg 1$ for scenarios with large interference and an ambitious SIR target give the desired result.

\subsection{Proof of Theorem \ref{th:Nvsm}}
The result (\ref{eqn:Ncon-1}) comes directly from (\ref{eqn:mg}) when substituting (\ref{eqn:op-ub2}). Now, if $\mu$ is reasonably small, then $\frac{k}{4}+\frac{1-\mu^2}{4\mu^2}$ will be large. As such, we can approximate $E_k(x)\approx\frac{e^{-x}}{x}$. We can therefore rewrite (\ref{eqn:Ncon-1}) as
\begin{align}
1+\sum_{k=1}^{N-1}\frac{{N-1 \choose k}(-1)^{k}}{\left(\frac{\sigma_{\rm I}^2\gamma}{\sigma^2}\right)^k}\left(\frac{1-\mu^2}{(k-1)\mu^2+1}\right)&\le 1-\frac{m}{N_{\rm I}+1},\notag\\
\Leftrightarrow ~\sum_{k=1}^{N-1}\frac{{N-1 \choose k}(-1)^{k+1}}{\left(\frac{\sigma_{\rm I}^2\gamma}{\sigma^2}\right)^k}\left(\frac{1-\mu^2}{1+(k-1)\mu^2}\right)&\ge\frac{m}{N_{\rm I}+1}.\label{eqn:d-1}
\end{align}
Noting that we have the Taylor series
\begin{equation}
\frac{1-\mu^2}{1+(k-1)\mu^2}=1-k\mu^2+(k-1)k\mu^4-(k-1)^2k\mu^6+\cdots\approx 1-k\mu^2,~\mbox{for small }\mu,
\end{equation}
we apply this approximation into (\ref{eqn:d-1}) to reach (\ref{eqn:Ncon-2}). The condition (\ref{eqn:Ncon-3}) can also be easily obtained by keeping only the first term in the summation which is valid if $\frac{\sigma_{\rm I}^2\gamma}{\sigma^2}$ is very large.

\subsection{Proof of Theorem \ref{th:W}}
To prove the result, we interpret the fluid antenna system as a linear space of $W\lambda$ that has $N$ ports having autocorrelation parameters in decreasing order, i.e.,
\begin{equation}\label{eqn:mu-decrease}
|\mu_2|\ge|\mu_3|\ge \cdots\ge |\mu_N|,
\end{equation}
as the position of the port moves away from the reference position, i.e., port $1$. The property in (\ref{eqn:mu-decrease}) does not normally apply because Bessel function $J_0(\cdot)$ is an oscillating function as the distance increases. However, as the distance from the reference position increases, the general trend of $|J_0(\cdot)|$ does go down. By defining the inverse of $J_0(\cdot)$ as the function $\rho^*=J_0^{-1}(\mu^*)$ that always returns the minimum value of $\rho^*$ to ensure $|J_0(\rho)|\le\mu^*$ for $\rho\ge\rho^*$ and enforce the monotonicity, we construct a fluid antenna model where the ports further away contribute more diversity than the near ones (which agrees with the intuition).

Based on this model, the performance of an $N$-port fluid antenna system is lower bounded by that of the same system, keeping only ports $\frac{N}{2}+1,\dots,N$ but dropping the ports from $1$ to $\frac{N}{2}$. This performance is further lower bounded by setting the autocorrelation parameters as the one at distance of $\frac{W\lambda}{2}$, i.e.,
\begin{equation}
\left|\mu_{\frac{N}{2}+1}\right|=\left|\mu_{\frac{N}{2}+2}\right|=\cdots=\left|\mu_{N}\right|=\left|J_0\left(\frac{2\pi}{\lambda}\frac{W\lambda}{2}\right)\right|=|J_0(\pi W)|.
\end{equation}
Setting $|J_0(\pi W)|=\mu^*$ in (\ref{eqn:mucon}) with these $\frac{N}{2}$ ports, and using $J_0^{-1}(\mu^*)$ provide a sufficient condition for $W$ in order to achieve the SIR target and multiplexing gain, which gives the desired result.

{\renewcommand{\baselinestretch}{1.1}
\begin{footnotesize}

\end{footnotesize}}

\end{document}

%% file: macro.tex
    \def\Complex{{\rm\rule[.23ex]{.03em}{1.1ex}\kern-.3em{C}}}

    \newcommand{\be}{\begin{equation}} \newcommand{\ee}{\end{equation}}
    \newcommand{\bea}{\begin{eqnarray}} \newcommand{\eea}{\end{eqnarray}}
    \newcommand{\benum}{\begin{enumerate}} \newcommand{\eenum}{\end{enumerate}}
    \def\proof{\noindent\hspace{2em}{\itshape Proof: }}
    
    \def\endproof{\hspace*{\fill}~$\square$\par\endtrivlist\unskip}

  \newtheorem{theorem}{Theorem}
  \newtheorem{definition}{Definition}
  
  \newtheorem{lemma}{Lemma}